\documentstyle[psfig]{mn}
\newcommand{\rsun}{$\mathrm{R_{\sun}}$}
\newcommand{\msun}{$\mathrm{M_{\sun}}$}
\newcommand{\lsun}{$\mathrm{L_{\sun}}$}
\newcommand{\mdot}{$\dot{M}$}
\newcommand{\mext}[2]{$\dot{M}_{\mathrm{#1},#2}$}
\newcommand{\rate}{$\mathrm{M_{\sun}} \, \mathrm{yr}^{-1}$}
\newcommand{\teff}{$T_{\mathrm{eff}}$}
\newcommand{\xpn}[2]{$#1 \times 10^{#2}$}

\title[The evolution of helium stars in close binary systems]
      {The evolution of naked helium stars \\ with a neutron-star companion 
       in close binary systems}
\author[Dewi et al.]
       {J. D. M. Dewi$^{1,3,4}$\thanks{email: jasinta@astro.uva.nl}, 
        O. R. Pols$^2$\thanks{current address: Astronomical Institute, 
                              University of Utrecht, Postbus 80000, 3508 TA Utrecht, 
                              The Netherlands. email: O.R.Pols@astro.uu.nl}, 
        G. J. Savonije$^1$, E. P. J. van den Heuvel$^1$\\
        $^1$Astronomical Institute {\it Anton Pannekoek},
            University of Amsterdam, Kruislaan 403, NL-1098 SJ Amsterdam,
            The Netherlands\\
        $^2$Department of Mathematics, P.O. Box 28M, Monash University, Clayton, 
            Victoria 3800, Australia\\
        $^3$Bosscha Observatory, Lembang 40391, Bandung Indonesia\\
        $^4$Department of Astronomy, Institut Teknologi Bandung, 
            Jl. Ganesha 10, Bandung 40132, Indonesia}
\date{Accepted . Received ; in original form }

\pagerange{\pageref{firstpage}--\pageref{lastpage}}
\pubyear{}

\begin{document}

\maketitle

\label{firstpage}

\begin{abstract}
The evolution of helium stars with masses of 1.5 -- 6.7~\msun\ in binary systems with a 1.4~\msun\ neutron-star companion is presented. Such systems are assumed to be the remnants of Be/X-ray binaries with B-star masses in the range of 8 -- 20~\msun\ which underwent a case B or case C mass transfer and survived the common-envelope and spiral-in process. The orbital period is chosen such that the helium star fills its Roche lobe before the ignition of carbon in the centre. We distinguish case BA (in which mass transfer is initiated during helium core burning) from case BB (onset of Roche-lobe overflow occurs after helium core burning is terminated, but before the ignition of carbon). We found that the remnants of case BA mass transfer from 1.5 -- 2.9~\msun\ helium stars are heavy CO white dwarfs. This implies that a star initially as massive as 12~\msun\ is able to become a white dwarf. CO white dwarfs are also produced from case BB mass transfer from 1.5 -- 1.8~\msun\ helium stars, while ONe white dwarfs are formed from 2.1 -- 2.5~\msun helium stars. Case BB mass transfer from more-massive helium stars with a neutron-star companion will produce a double neutron-star binary. We are able to distinguish the progenitors of type Ib supernovae (as the high-mass helium stars or systems in wide orbits) from those of type Ic supernovae (as the lower-mass helium stars or systems in close orbits). Finally, we derive a "zone of avoidance" in the helium star mass vs. initial orbital period diagram for producing neutron stars from helium stars.
\end{abstract}



\section{Introduction}
\label{helium:sec:intro}

A binary pulsar with a neutron-star or a heavy white-dwarf companion has long been considered to originate from a helium star in a binary system with a neutron-star companion. The latter system is a descendant of a high-mass X-ray binary (HMXB), in which the companion of the neutron star loses its mass through wind mass loss or a mass-transfer phase, exposing its helium core (e.g. Bhattacharya \& van den Heuvel 1991). Although the existence of such a system is only found in Cyg X-3 (van Kerkwijk et al. 1992, 1996), a detailed study of the evolution of helium stars in binary systems with a compact companion is important as the systems form the bridge between the evolution of X-ray binaries and the formation of double compact-object binaries.

The evolution of a helium star in a binary system has been studied e.g. by Savonije, de Kool \& van den Heuvel (1986) who evolved a 0.6~\msun\ non-degenerate helium star with a 1.3~\msun\ compact companion. With a short orbital period of $P = 37^{\mathrm{m}}$, Roche-lobe overflow (RLOF) takes place during helium core burning. Ergma \& Fedorova (1990) evolved helium stars with masses of 0.5, 0.766, and 1~\msun. As the companion star, they took white dwarfs with the same range of masses in a combination such that the systems have a mass ratio of $0.5 \leq M_{\mathrm{He}}/M_{\mathrm{WD}} \leq 2$. The periods are also so short, of $26\fm2$ -- $62\fm6$, that the helium stars transfer mass to the companion while they are still burning helium in the centre. A similar work was also carried out by Tutukov \& Fedorova (1990). A study of systems with more massive helium stars in wider orbits has been carried out by e.g. Delgado \& Thomas (1981) who considered helium stars with masses of 2, 2.7, 3.3, and 4~\msun\ with a massive main-sequence companion. The helium star fills its Roche lobe after helium is exhausted in the core. Habets (1986a) evolved a 2.5~\msun\ helium star with a 17~\msun\ main-sequence companion in a wide orbit ($P = 20\fd25$) such that mass-transfer phase occurs during first carbon burning convective shell.

In this work we study the evolution of helium stars with masses of 1.5 -- 6.7~\msun\ with a 1.4~\msun\ neutron-star companion. Such systems are assumed to be the remnants of Be/X-ray binaries with masses in the range of 8 -- 20~\msun\ which underwent mass transfer as case B (RLOF is initiated during hydrogen shell burning) or case C (during helium shell burning). As the result of the large mass ratio, mass transfer from the Be star to the neutron star is dynamically unstable and the two components are embedded in a common envelope (CE), leading to the spiraling-in of the neutron star in the envelope of the Be star. We assume that the system survived the common envelope and spiral-in process. We are interested to investigate the final fate of the systems (whether they will become white-dwarf/neutron-star or double neutron-star binaries), the type of supernovae (SN) they might produce, as well as to study which systems are stable to RLOF. A study on the similar range of mass (2 -- 6~\msun\ helium stars) has been carried out by Avila Reese (1993). With a Roche radius of 0.6 and 0.7~\rsun, in that study RLOF takes place after helium core burning is terminated and the calculations were done up to the ignition of carbon in the centre. Apart from the slightly wider range of mass in our study compared to the latter work and a larger range of orbital periods used in our work, we also are able to follow the evolution of the helium stars to more advanced evolutionary stage, i.e. beyond the carbon ignition, which enables more detailed conclusions on the fate of the systems. We take 6.7~\msun\ as the upper limit of our calculation because more massive helium stars undergo a dynamically unstable mass transfer.

In Sect.~\ref{helium:sec:evolution} we describe the possibility for the formation channel of a helium star in a binary system with a neutron-star companion, and some basic assumptions used in the calculation of the orbital evolution. The computational code, the input parameters used in the code, and the calculation that constrains the initial mass of the helium star and the initial period are described in Sect.~\ref{helium:sec:method}. We discuss the results in two sections; Sect.~\ref{helium:sec:BA} for the cases in which RLOF takes place during helium core burning and Sect.~\ref{helium:sec:BB} for that during helium shell burning. Our conclusions are given in Sect.~\ref{helium:sec:conclusion}.


\section{The evolution of helium star in binary system}
\label{helium:sec:evolution}


\subsection{Formation of a helium star with a neutron-star companion in a binary 
system}
\label{helium:subsec:formation}

An HMXB consists of a massive OB-type star and a compact object as the X-ray source. There are two classes of HMXB, i.e.: (i) OB-supergiant (standard massive) X-ray binaries, in which the X-ray source is powered by wind accretion from its almost Roche-lobe filling OB-supergiant companion, and (ii) Be/X-ray binaries, in which the X-ray source is powered by accretion of material when the compact object moves through the dense disc of its unevolved Be-star companion at periastron passage. The first class is found in close, almost circular orbits with companion stars more massive than 20~\msun; while the latter in wide, eccentric orbits, with 8 -- 20~\msun\ companion stars.

As the companion starts to fill its Roche lobe, a runaway mass transfer to the compact object will follow due to the high mass ratio of the system, leading to a CE and spiral-in phase. The outcome of this phase is a system consisting of the helium core of the companion and a neutron star. The helium star evolves and, in turn, might fill its Roche lobe. If the helium star is more massive than 2.2~\msun\ (Habets 1986a) and it does not further lose mass, it will collapse and produce a neutron star; and the system becomes a double neutron-star binary. We refer to Bhattacharya \& van den Heuvel (1991) for review on the so-called "standard scenario"\footnote{An evolutionary channel for the formation of double neutron-star binary similar to the standard scenario was also proposed by Tutukov \& Yungelson (1973). They predicted a helium-star mass of 2~\msun\ as the limiting mass for white-dwarf/neutron-star formation.}.

An alternative to the standard model was proposed by Brown (1995). This new evolutionary scenario suggests that a double neutron-star binary is a descendant of a double helium-star binary, which is produced from a binary system with components of very similar masses. This twin-mass model was proposed because of the argument that a neutron star would be able to accrete matter at high rate during the CE and spiral-in process and will then become a black hole (e.g. Chevalier 1993), which would exclude the possibility that double neutron stars would result from the CE evolution of HMXB (the standard scenario). Nevertheless, whether or not a neutron star in a CE will indeed go to highly super-Eddington accretion is still very uncertain (see e.g. Chevalier 1996). Therefore, there is still room for the formation of systems consisting of a helium star with a neutron-star companion like in the standard scenario. Strong accretion onto the neutron star can be avoided, for example, if the donor star is very extended (Chevalier 1993) and if rotational effects are taken into account (Chevalier 1996).


\subsection{Orbital evolution}
\label{helium:subsec:orbital}

The orbital angular momentum of a binary system which contains a helium star of mass $M_{\mathrm{He}}$ and a neutron star of mass $M_{\mathrm{NS}}$ is given by
	\begin{eqnarray}
	J^{2}_{\mathrm{orb}} & = & 
	\frac{M^{2}_{\mathrm{He}} M^{2}_{\mathrm{NS}}}{M} \,\, a \, G
	\label{helium:eq:angular}	 
	\end{eqnarray}
where $M = M_{\mathrm{He}} + M_{\mathrm{NS}}$ is the total mass of the system, $a$ is the orbital separation, and $G$ is the constant of gravity. The rate of change in orbital separation of the system is expressed by
	\begin{eqnarray}
        \frac{\dot{a}}{a} & = & 2 \, 
        \frac{\dot{J}_{\mathrm{orb}}}{J_{\mathrm{orb}}} 
        - 2 \, \left( \frac{\dot{M}_{\mathrm{He}}}{M_{\mathrm{He}}} + 
        \frac{\dot{M}_{\mathrm{NS}}}{M_{\mathrm{NS}}} \right) + 
        \frac{\dot{M}}{M}
	\label{helium:eq:separation}	 
	\end{eqnarray}
The total change in orbital angular momentum is assumed to be affected by gravitational-wave radiation and by the loss of mass with angular momentum from the system, i.e.
	\begin{eqnarray}
        \frac{\dot{J}_{\mathrm{orb}}}{J_{\mathrm{orb}}} & = &  
        \frac{\dot{J}_{\mathrm{gwr}}}{J_{\mathrm{orb}}} + 
        \frac{\dot{J}_{\mathrm{ml}}}{J_{\mathrm{orb}}}
	\label{helium:eq:total}	 
	\end{eqnarray}
where the rate of change of orbital angular momentum due to gravitational-wave radiation is given by (Landau \& Lifshitz 1958)
	\begin{eqnarray}
        \frac{\dot{J}_{\mathrm{gwr}}}{J_{\mathrm{orb}}} & = & - 
        \frac{32 \, G^{3}}{5 \, c^{5}} \,\,  
        \frac{M_{\mathrm{He}} M_{\mathrm{NS}} M}{a^{4}}
	\label{helium:eq:wave}	 
	\end{eqnarray}
with $c$ the speed of light in vacuum.

The system is assumed to evolve non-conservatively. Mass loss from the system causes loss of orbital angular momentum which is taken into account as (van den Heuvel 1994; Soberman, Phinney \& van den Heuvel 1997)
	\begin{eqnarray}
        \frac{\dot{J}_{\mathrm{ml}}}{J_{\mathrm{orb}}} & = & 
        \frac{\alpha + \beta q^{2}}{1 + q} \,\,
        \frac{\dot{M}_{\mathrm{He}}}{M_{\mathrm{He}}}
	\label{helium:eq:massloss}	 
	\end{eqnarray}
where $q = M_{\mathrm{He}} / M_{\mathrm{NS}}$ is the mass ratio; $\alpha$ is the fraction of mass lost from the helium star in the form of fast isotropic wind, and $\beta$ is the fraction of mass ejected isotropically from the vicinity of the neutron star. During the detached phase, we took $\alpha = 1$ and it is assumed that the neutron star does not accrete matter from the stellar wind ($\beta = 0$); all matter released from the system in this case carries the specific orbital angular momentum of the helium star. The stellar wind mass loss is approximated by eq. (2) in Wellstein \& Langer (1999, cf. Hamann, Sch\"{o}nberner \& Heber 1982; Hamann, Koesterke \& Wessolowski 1995) multiplied by a factor of 0.5 to provide a better fit with the most recent observed mass-loss rate of Wolf-Rayet stars (Nugis \& Lamers 2000; Nelemans \& van den Heuvel 2001), i.e.
	\begin{eqnarray}
        \dot{M}_{\mathrm{He,wind}} & = & 
        \left\{ \begin{array}{ll} 
                2.8 \, 10^{-13} (L/\mathrm{L_{\sun}})^{1.5}, & 
                \log{L/\mathrm{L_{\sun}}} \geq 4.5 \\
                4.0 \, 10^{-37} (L/\mathrm{L_{\sun}})^{6.8}, & 
                \log{L/\mathrm{L_{\sun}}} < 4.5 
               \end{array} \right.
	\label{helium:eq:wind}	 
	\end{eqnarray}
This means our adopted mass loss rate in the upper luminosity range is one quarter of the Hamann et al. (1995) rate. During RLOF, the helium star transfers mass with a mass-transfer rate which is defined as
	\begin{eqnarray}
        \dot{M}_{\mathrm{He,RLOF}} & = & 
        - 10^{3} \times \mathrm{max}\left[0, 	
        \left(\ln{\frac{R}{R_{\mathrm{L}}}}\right)^{3}\right]
	\label{helium:eq:rlof}	 
	\end{eqnarray}
where $R_{\mathrm{L}}$ is the helium star's Roche radius given by Eggleton (1983) as
	\begin{eqnarray}
        R_{\mathrm{L}} & = & \frac{0.49 \,\, a}{0.6 + q^{-2/3} 
        \ln{(1 + q^{1/3})}}
	\label{helium:eq:roche}	 
	\end{eqnarray}
During the mass-transfer phase (even on a nuclear timescale) the wind mass loss contributes at most 5 per cent of the mass-loss rate. It is assumed that the neutron star accretes matter up to its Eddington limit for helium accretion (\xpn{3}{-8} \rate); the rest of the transferred matter is lost from the system with the specific orbital angular momentum of the neutron star.

In the limiting case when the mass-transfer rate owing to RLOF is much higher than both the maximum accretion rate and the wind mass-loss rate (\mdot\ $> 10^{-6}$~\rate), we have $\beta \approx 1$ and $\dot{J}_{\mathrm{gwr}}$ can be neglected. This is generally the case during case BAB and case BB mass transfer discussed below. Then eq.~(\ref{helium:eq:massloss}) leads to:

	\begin{eqnarray}
	\frac{\dot{P}}{P} = \frac{3q^2 - 2q - 3}{1 + q} \,\,
	\frac{\dot{M}_{\mathrm{He}}}{M_{\mathrm{He}}}
	\label{helium:eq:pdot}	 
	\end{eqnarray}

From this it can be seen that $P$ has a minimum when $q = 1.39$, i.e. $M_{\mathrm{He}} \approx 1.95$. For $q > 1.39$, $P$ initially decreases as a result of RLOF, and increases again after $q < 1.39$. This behaviour can be seen in Sect.~\ref{helium:sec:BA} and \ref{helium:sec:BB}.


\section{The method of calculation}
\label{helium:sec:method}


\subsection{A brief description on the numerical code}
\label{helium:subsec:code}

We used the Eggleton stellar-evolution code (Eggleton 1971, 1972, 1973). This code uses a self-adaptive, non-Lagrangian mesh-spacing which is a function of local pressure, temperature, Lagrangian mass and radius. It treats both convective and semi-convective mixing as a diffusion process and finds a simultaneous and implicit solution of both the stellar structure equations and the diffusion equations for the chemical composition. The most recent update, including the nuclear reaction network, is described in Pols et al. (1995).

We evolved zero-age main-sequence stars (in Sect.~\ref{helium:subsec:initial}) assuming a chemical composition of (${X} = 0.70, {Z} = 0.02$) and using a mixing-length parameter of $\alpha = l/H_{p} = 2.0$. Convective overshooting is taken into account in the same way as in Pols~et~al. (1998) using an overshooting constant $\delta_{\mathrm{ov}} = 0.12$.

The helium stars were evolved assuming a chemical composition of (${Y} = 0.98, {Z} = 0.02$). All calculations were done without enhanced mixing (STD models in Pols 2002). We refer to the latter paper for the evolution of single helium stars related to our calculations.


\subsection{Initial conditions}
\label{helium:subsec:initial}

\subsubsection{Progenitor systems of helium star and neutron star binaries}
\label{helium:subsubsec:progenitor}

To obtain estimates of the mass of the helium star and the initial period, we evolved massive main-sequence stars with masses of 8 -- 20~\msun. If the radius of the star is assumed to be equivalent to its Roche radius when RLOF is initiated ($R \approx R_{\mathrm{L}}$), then the separation of the binary system at the onset of the mass-transfer phase, $a_{\mathrm{i}}$, for a given radius $R$ can be calculated with eq.~(\ref{helium:eq:roche}). With the energy equation for CE evolution (Webbink 1984; de Kool 1990)
	\begin{eqnarray}
	  \frac {a_{\mathrm{f}}} {a_{\mathrm{i}}} & = & 
	  \frac {M_{\mathrm{core}} M_{\mathrm{2}}} {M_{\mathrm{donor}}} \,\,
	  \frac {1} {M_{\mathrm{2}} + 2 M_{\mathrm{env}}/
	  (\eta_{\mathrm{CE}}\lambda r_{\mathrm{L}})} 
	  \label{helium:eq:webbink}
	\end{eqnarray}
we calculated the final separation, $a_{\mathrm{f}}$. Here $M_{\mathrm{donor}} = M_{\mathrm{core}} + M_{\mathrm{env}}$, $M_{\mathrm{core}}$, and $M_{\mathrm{env}}$ are the masses of the donor star, its helium core and envelope, respectively. Due to the stellar wind mass loss (de Jager, Nieuwenhuijzen \& van der Hucht 1988; Nieuwenhuijzen \& de Jager 1990), $M_{\mathrm{donor}}$ might be smaller than $M_{\mathrm{ZAMS}}$, the mass of the star on the zero-age main sequence (ZAMS). During the red supergiant phase, the stellar mass is reduced by 3 -- 19 per cent (in a 8~\msun\ and a 20~\msun\ stars, respectively). $M_{2}$ is the mass of the companion, which we assumed to be a 1.4~\msun\ neutron star. $r_{\mathrm{L}} = R_{\mathrm{L}} / a_{\mathrm{i}}$ is the dimensionless Roche radius of the donor star. We took $\eta_{\mathrm{CE}}$, the so-called CE efficiency parameter, to be one. $\lambda$, the parameter of the binding energy of the envelope to the core, is approximated as in Dewi \& Tauris (2000). We assumed that the binary will survive the CE and spiral-in phase if the final separation is larger than the Roche radius of the helium core, i.e. the helium star does not immediately fill its Roche lobe. Following Taam, King \& Ritter's (2000) argument, we assume that the neutron star does not accrete a significant amount of matter while it is embedded in the donor's envelope.

	\begin{figure}
 	 \centerline{\psfig{file=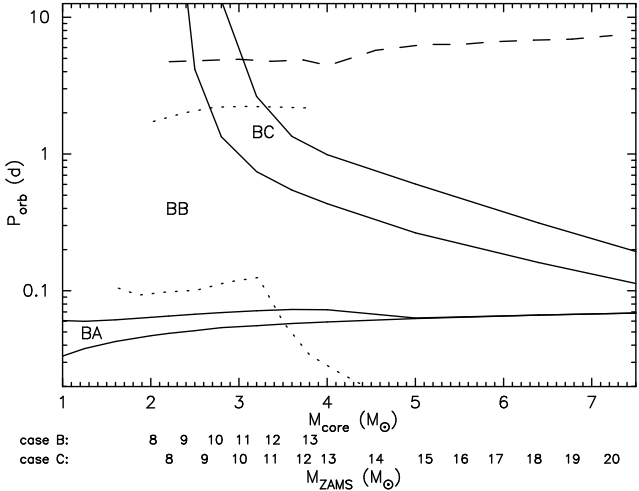,width=\linewidth}}
	 \caption[]{Mass of the helium core and orbital period following a CE 
	  evolution resulting from case B and C mass transfers from 8 -- 
	  20~\msun\ main-sequence stars. The numbers below the figure indicate 
	  the corresponding initial ZAMS mass. The full lines separate the 
	  regions of case BA, BB, and BC mass transfers, taken from Pols' (2002) 
	  single helium-star calculations. The dotted lines mark the maximum 
	  period allowed resulting from case B (lower- and upper-line, derived 
          from the two extremes of $\lambda$-value); and the dashed line 
          resulting from case C mass transfer. See text for an explanation.}
	 \label{helium:fig:initial}
	\end{figure}

The results of the calculations are summarized in Fig.~\ref{helium:fig:initial}, where we have plotted the maximum periods (which are related to the maximum radius) after case B and case C spiral-in. These values are very uncertain, especially for case B, because the parameter $\lambda$ used for calculating the post-CE separation depends quite sensitively on the choice of the mass of the helium core. As argued by Tauris \& Dewi (2001) approximate lower limits to $\lambda$ and to the core mass are obtained by assuming that the core mass is the mass where the abundance of hydrogen is below 0.1. Considering the bottom of convective envelope as the core, results in approximate upper limits of $\lambda$ and the core mass. The latter criterion corresponds roughly to the mass where $X_{\mathrm{H}} = 0.3$, for the maximum radius during case B from $M_{\mathrm{ZAMS}} \leq 13$~\msun. The dotted lines in Fig.~\ref{helium:fig:initial} mark the maximum periods for case B evolution, calculated with the two extreme $\lambda$-values. For case C evolution, only the maximum period (dashed-line) corresponding to the lower limit of $\lambda$ is plotted. In stars with $M_{\mathrm{ZAMS}} \ga 14$~\msun, helium is ignited before a deep convective envelope has developed, and the radius keeps expanding during core He burning. At the maximum radius for case B (which in this case is not well-defined), the envelope is more strongly bound and the maximum period after case B is correspondingly smaller (see Fig.~\ref{helium:fig:initial}). In this case, the bottom of the convective envelope can no longer be realistically taken as the core mass, so the upper dotted line in Fig.~\ref{helium:fig:initial} stops at 13~\msun. Because of these uncertainties, to relax the assumption on the choice of core mass as well as on $\eta_{\mathrm{CE}}$ (i.e. the possibility of applying $\eta_{\mathrm{CE}}$ other than unity), we did the calculations for a wide range of periods.

\subsubsection{Studied types of helium star binary evolution}
\label{helium:subsubsec:cases}

\setlength{\tabcolsep}{1.5pt}
        \begin{table}
         \caption[]{Terminology of the mass-transfer phases}
         \label{helium:table:types}
         \begin{center}
         \begin{tabular}{lll}
          \hline
          \hline
          \noalign{\smallskip}
                                         & Case B          & Case C              \\
          \noalign{\smallskip}
          \hline
          \hline
          \noalign{\smallskip}
          Occurence                      & H-shell burning & He-shell burning    \\
          Outcome                        & He star         & He star $+$ CO-core \\
          \noalign{\smallskip}
          \hline
          \noalign{\smallskip}
          \multicolumn{3}{l}{{\bf Helium star evolution}}                        \\
          RLOF during                    &                 &                     \\
          $\bullet$ He-core burning      & Case BA         &                     \\
          ~~ successive He-shell burning & Case BAB        &                     \\
          $\bullet$ He-shell burning     & Case BB         & Case CB             \\
          $\bullet$ C-burning            & Case BC         & Case CC             \\
          \noalign{\smallskip}
          \hline
         \end{tabular}
         \end{center}
         \end{table}
\setlength{\tabcolsep}{6.0pt}

As Fig.~\ref{helium:fig:initial} shows, the progenitors of helium star and neutron star binaries were main-sequence star and neutron star systems that evolved according to case B or case C mass transfer. The outcome of CE evolution in case B is an unevolved helium star and neutron star, while that in case C is an evolved helium star (which has developed a carbon core) and neutron star. The systems resulting from case B can be so close that they start RLOF during helium core burning. This is so-called case BA evolution. Wider systems will start RLOF only during helium shell burning around the carbon core. This is so-called case BB evolution. The terminology of the types of evolution used in this paper is summarized in Table~\ref{helium:table:types}.

Case C mass transfer will result in evolved helium stars which, after CE evolution, can only start transferring mass after helium core burning is terminated. We are interested in RLOF during helium shell burning, which can be called case CB evolution. Basically, case BB and CB evolutions result in the same type of RLOF from a helium star of about the same mass and period, except in $M_{\mathrm{He}} \geq 4$~\msun\ where the wind mass loss plays a major role in reducing the mass and changing the structure during helium core burning. We do not separately study case CB evolution in this paper, but assume that it is similar to case BB evolution.

A single helium star expands and reaches the maximum radius during convective carbon shell burning (Habets 1986a; Pols 2002). Although they are not less important, we did not attempt to follow evolutions in which RLOF is initiated during carbon core burning or beyond (case BC or CC evolution). The overall results for case BA mass transfer are shown in Table~\ref{helium:tab:caseBA}, and those for case BB in Tables~\ref{helium:tab:BBlowmass} and \ref{helium:tab:BBhighmass}.


\section{Results: Case BA Mass Transfer}
\label{helium:sec:BA}

	\begin{figure}
 	 \centerline{\psfig{file=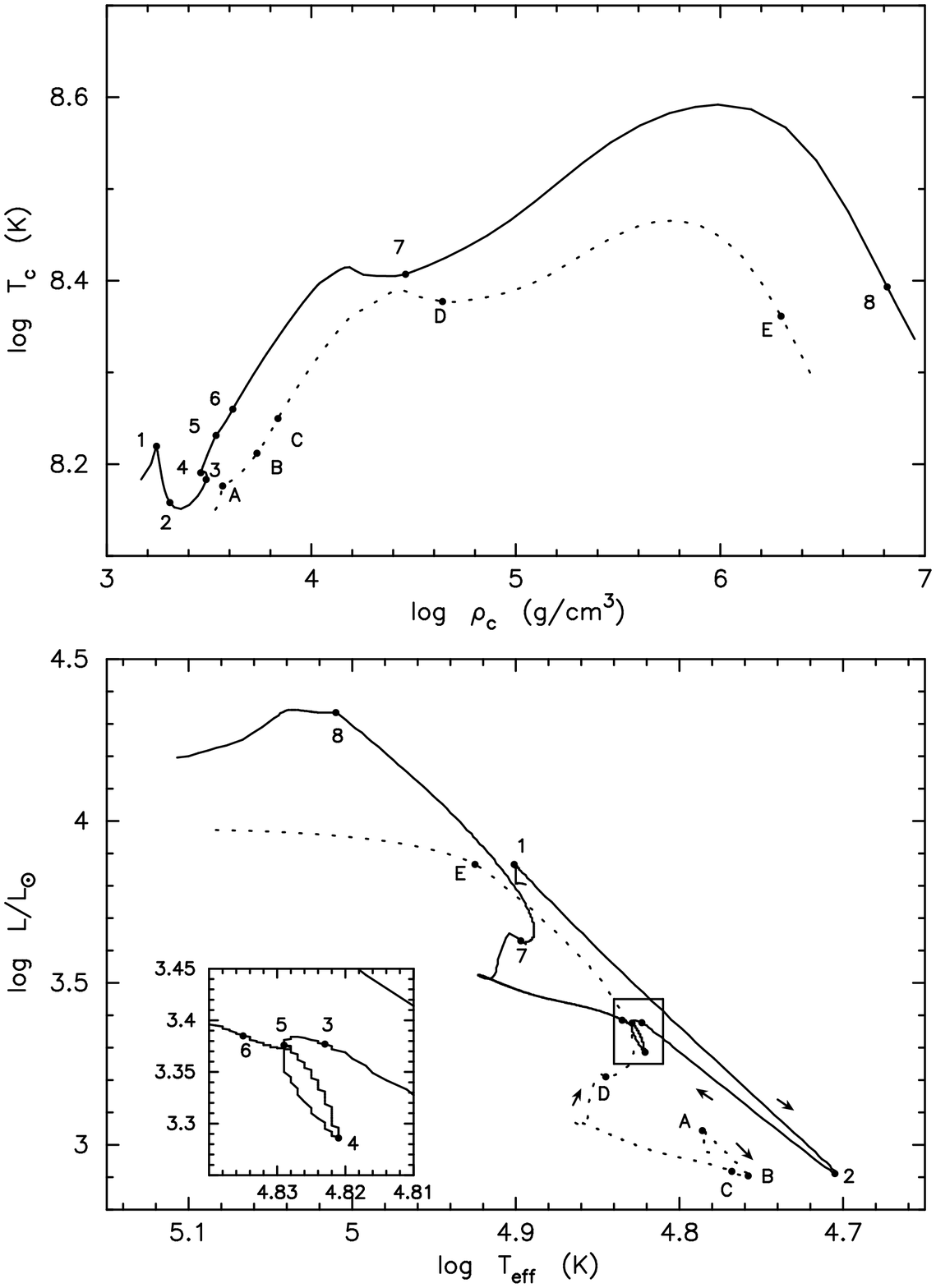,width=\linewidth}}
	 \caption[]{The evolutionary tracks in the HR diagram (bottom) and 
          central density-central temperature plane (top) for case BA mass 
          transfer from a 1.5~\msun\ (dotted-) and a 2.8~\msun\ helium stars 
	  (solid-line). See text for an explanation of the labels.}
	 \label{helium:fig:central}
	\end{figure}

	\begin{table*}  
         \caption[]{The binary parameters in case BA mass transfer from 1.5 -- 
          2.9~\msun\ helium stars: the initial mass and period; the duration, 
          amount of mass transferred from the helium star, and the maximum 
          mass-loss rate during the case BA mass transfer; the duration of the 
          detached phase; the duration, amount of mass transferred from the 
          helium star, and the maximum mass-loss rate during the case BAB mass 
          transfer; the final mass and period, the final mass of helium in the 
          envelope and the final mass of the neutron star. Masses are in solar 
          unit, mass-loss rates in \rate, times in yr and periods in days. 
          $^{*}$ denotes the slow phase of mass transfer.}
	 \label{helium:tab:caseBA}
	 \begin{center}
	 \begin{tabular}{r@{\ \ }rr@{\ \ }r@{\ \ }rrr@{\ \ }r@{\ \ }r
	                 r@{\ \ }r@{\ \ }r@{\ \ }r}
          \noalign{\smallskip}
	  \hline
	  \hline
          \noalign{\smallskip}
          $M_{\mathrm{i}}$ & $P_{\mathrm{i}}$~~ & \multicolumn{3}{c|}{Case BA} &   
          $\Delta t_{\mathrm{detached}}$ & \multicolumn{3}{c|}{Case BAB} & 
          $M_{\mathrm{f}}$~ & $P_{\mathrm{f}}$~ & $M_{\mathrm{He,e}}$ & 
          $M_{\mathrm{NS,f}}$ \\
          & & $\Delta t_{\mathrm{RLOF}}$~ & $\Delta M$ & 
          $\dot{M}_{\mathrm{max}}$~~~ & & $\Delta t_{\mathrm{RLOF}}$~~ & 
          $\Delta M$ & $\dot{M}_{\mathrm{max}}$~~~ &  & & \\
          \noalign{\smallskip}
	  \hline
	  \hline
          \noalign{\smallskip}
          1.5 & 0.050  & 
                \xpn{4.51}{6} & 0.290 & \xpn{1.4}{-7} & \xpn{1.17}{6} & 
                \xpn{8.27}{5} & 0.425 & \xpn{1.4}{-6} & 0.785 & 0.11 & 
                0.0350 & 1.544 \\
              & 0.060  & 
                \xpn{9.69}{5} & 0.042 & \xpn{8.3}{-8} & \xpn{9.88}{5} & 
                \xpn{4.96}{5} & 0.580 & \xpn{3.1}{-6} & 0.864 & 0.12 & 
                0.0262 & 1.439 \\
          1.8 & 0.050  & 
                \xpn{3.84}{6} & 0.440 & \xpn{3.3}{-7} & \xpn{9.94}{5} & 
                \xpn{6.00}{5} & 0.513 & \xpn{2.2}{-6} & 0.836 & 0.12 & 
                0.0276 & 1.523 \\
              & 0.060  & 
                \xpn{1.08}{6} & 0.138 & \xpn{2.7}{-7} & \xpn{8.76}{5} & 
                \xpn{3.48}{5} & 0.715 & \xpn{5.2}{-6} & 0.933 & 0.13 & 
                0.0168 & 1.441 \\
          2.1 & 0.050  & 
                \xpn{3.74}{6} & 0.683 & \xpn{1.2}{-6} & \xpn{9.81}{5} & 
                \xpn{5.41}{5} & 0.556 & \xpn{2.7}{-6} & 0.857 & 0.12 & 
                0.0277 & 1.521 \\
              & 0.060  & 
                \xpn{1.16}{6} & 0.360 & \xpn{9.7}{-7} & \xpn{8.07}{5} & 
                \xpn{3.03}{5} & 0.768 & \xpn{6.1}{-6} & 0.970 & 0.13 & 
                0.0133 & 1.443 \\
          \noalign{\smallskip}
	  \hline
          \noalign{\smallskip}
          2.4 & 0.060  & 
                \xpn{1.37}{6} & 0.681 & \xpn{5.0}{-6} & \xpn{8.13}{5} & 
                \xpn{3.15}{5} & 0.731 & \xpn{5.4}{-6} & 0.981 & 0.12 & 
                0.0119 & 1.449 \\
              & $^{*}$ & 
                              &       & \xpn{1.1}{-6} &               &          
                              &       &               &       &      &
                       &       \\
              & 0.065  & 
                \xpn{5.16}{5} & 0.432 & \xpn{4.6}{-6} & \xpn{6.25}{5} & 
                \xpn{1.92}{5} & 0.877 & \xpn{9.8}{-6} & 1.089 & 0.11 & 
                0.0147 & 1.421 \\
              & $^{*}$ & 
                              &       & \xpn{8.6}{-7} &               &          
                              &       &               &       &      & 
                       &       \\
          2.5 & 0.060  & 
                \xpn{1.48}{6} & 0.806 & \xpn{9.5}{-6} & \xpn{8.29}{5} & 
                \xpn{3.32}{5} & 0.707 & \xpn{5.1}{-6} & 0.976 & 0.12 & 
                0.0122 & 1.453 \\
              & $^{*}$ & 
                              &       & \xpn{9.9}{-7} &               &           
                              &       &               &       &      & 
                       &       \\
              & 0.065  & 
                \xpn{6.58}{5} & 0.585 & \xpn{9.8}{-6} & \xpn{6.17}{5} & 
                \xpn{2.09}{5} & 0.821 & \xpn{8.1}{-6} & 1.079 & 0.11 & 
                0.0109 & 1.425 \\
              & $^{*}$ &
                              &       & \xpn{7.2}{-7} &               &          
                              &       &               &       &      & 
                       &       \\
          \noalign{\smallskip}
	  \hline
          \noalign{\smallskip}
          2.8 & 0.060  & 
                \xpn{5.50}{4} & 1.061 & \xpn{4.9}{-5} & \xpn{1.45}{6} & 
                \xpn{3.47}{5} & 0.713 & \xpn{5.4}{-6} & 0.968 & 0.12 & 
                0.0091 & 1.429 \\
	      & $^{*}$ & 
	        \xpn{6.16}{5} & 0.049 & \xpn{1.5}{-7} & \xpn{7.90}{5} &          
	                      &       &               &       &      &  
	               &       \\
              & 0.065  & 
                \xpn{5.51}{4} & 1.032 & \xpn{5.1}{-5} & \xpn{1.74}{6} & 
                \xpn{2.81}{5} & 0.721 & \xpn{5.3}{-6} & 1.041 & 0.11 & 
                0.0057 & 1.410 \\
          2.9 & 0.060  &  
                \xpn{5.56}{4} & 1.394 & \xpn{7.3}{-5} & \xpn{3.92}{6} & 
                \xpn{4.38}{5} & 0.577 & \xpn{2.4}{-6} & 0.922 & 0.13 & 
                0.0047 & 1.415 \\
              & 0.065  & 
                \xpn{5.43}{4} & 1.330 & \xpn{7.6}{-5} & \xpn{2.29}{6} & 
                \xpn{3.49}{5} & 0.566 & \xpn{2.0}{-6} & 0.999 & 0.12 & 
                0.0071 & 1.412 \\
          \noalign{\smallskip}
	  \hline
	  \hline
         \end{tabular}
	 \end{center}
	\end{table*}

Based on the stability of RLOF during helium core burning, we devide case BA mass transfer into three groups, i.e. (i) 1.5 -- 2.1~\msun\ helium stars, in which RLOF is stable, (ii) 2.4 -- 2.5~\msun\ helium stars, in which RLOF is initiated by a rapid, unstable phase and followed by a slow, stable phase of mass transfer, and (iii) 2.8 -- 2.9~\msun\ helium stars, in which the unstable, thermal-timescale RLOF causes the temporary disappearance of the convective core. 

In general, case BA mass transfer from helium stars in our calculations resembles case A evolution from main-sequence stars (e.g. Pols 1994; Wellstein, Langer \& Braun 2001). In all helium stars of 1.5 -- 2.9~\msun\ which undergo case BA mass transfer, the whole envelope is removed from the star, and the remnant, a degenerate CO white dwarf, is never massive enough to ignite carbon.


\subsection{Roche-lobe overflow from 1.5 -- 2.1 $\bmath{\mathrm{M_{\sun}}}$ 
helium stars}
\label{helium:subsec:BA1p5}

\begin{figure}
 	 \centerline{\psfig{file=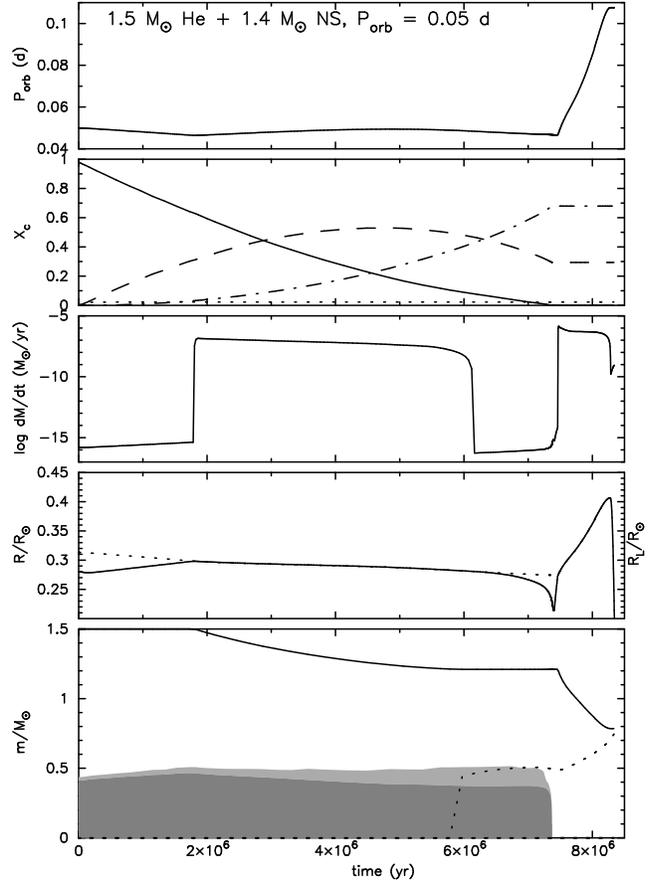,width=\linewidth}}
	 \caption[]{{\bf Case BA.} The evolution of a 1.5~\msun\ helium star with 
          a 1.4~\msun\ neutron-star companion in a very close orbit. The upper 
          panel shows the orbital evolution. The second panel shows the central 
          abundances. Solid-, dashed-, dash-dotted-, and dotted-lines are the 
          helium, carbon, oxygen, and neon abundances, respectively. The third 
          panel gives the mass-loss rate in \rate; wind mass-loss rate during 
          the detached phase, or mass-transfer rate during the Roche-lobe 
          overflow phase. The fourth panel presents the stellar (solid-) and 
          Roche radii (dotted-line), in solar radius. The fifth panel shows the 
          evolution of the stellar interior. The solid line represents the total 
          mass of the donor star; the dotted line the CO core mass (defined as 
          the mass where the helium abundance is below 0.1). The dark- and light-
          shaded areas mark the helium-burning convective and semiconvective 
          regions.}
	 \label{helium:fig:BA1p5}
	\end{figure}

During the detached phase the orbit slowly shrinks due to gravitational-wave radiation and the stellar radius slowly increases, until (stable) RLOF starts (point A in Fig.~\ref{helium:fig:central}). During the first phase of stable, nuclear-timescale RLOF, the star contracts and the energy-production rate in the centre decreases as a result of the loss of mass, which causes the luminosity and effective temperature to decrease. The star remains in thermal equilibrium. The mass of the convective core does not grow as in the case of a single star, but decreases slowly until the core disappears when the helium is exhausted (see Fig.~\ref{helium:fig:BA1p5}). Most of the transferred matter is lost from the system at low specific orbital angular momentum which, although counteracted by gravitational-wave radiation, causes the orbit to slightly expand. Roughly at the same time when the abundance of carbon in the centre is maximum, the period reaches a maximum value, while $L$ and \teff\ reach their minimum values (the reddest point in the HR diagram, point B). After this point, the orbit gradually shrinks. We find that the earlier RLOF is initiated (the smaller the initial period), the more mass is removed from the helium star, the more significant the change in period, and the longer the duration of RLOF. 

After the turning point, the star moves to the left in the HR diagram, only an insignificant amount of mass is transferred from the helium star -- this is also the case if the orbital period is such that mass transfer starts after the carbon abundance in the centre reaches its maximum. In the latter case, there is no maximum period; the period continuously decreases (while the star shrinks) until the system is detached. The maximum mass-loss rate, the duration of RLOF, and the amount of mass transferred to the neutron star all decrease with decreasing initial mass ratio and with increasing initial period.

As the helium abundance in the core drops below about 0.1, the star contracts inside its Roche lobe, and the system becomes detached (point C). During this detached phase, the orbit is again tightened due to gravitational-wave radiation, but the net change in orbit is insignificant compared to the initial period. The duration of helium core burning is prolonged due to the mass transfer. However, we found that if the initiation of RLOF takes place close to the end of the helium main sequence, helium core burning proceeds marginally faster (by less than 1 per cent) than in a single helium star of the same mass.

After helium is exhausted in the centre and the convective core disappears, helium burning moves to a radiative shell around the CO core. The CO core grows in mass but decreases in radius, which causes the star to expand and move up in the HR diagram. A second phase of stable RLOF, which we call case BAB, then takes place (point D). The central temperature increases. The mass-loss rate, which is higher than that during the helium core burning, increases both with  initial mass ratio and with initial period. The orbit expands in all cases due to the effects described in Sect.~\ref{helium:subsec:orbital}. At the end of this mass-transfer phase, when almost the entire envelope is removed from the star, the degeneracy border is crossed. The central temperature decreases as well as the energy-production rate. Although the core is constant in mass, the radius decreases and the star contracts inside its Roche lobe. The system is again detached (point E). The duration of the second RLOF phase decreases with increasing initial period and initial mass ratio.

\subsection{Roche-lobe overflow from 2.4 -- 2.5 $\bmath{\mathrm{M_{\sun}}}$ 
helium stars}
\label{helium:subsec:BA2p5}

\begin{figure}
 	 \centerline{\psfig{file=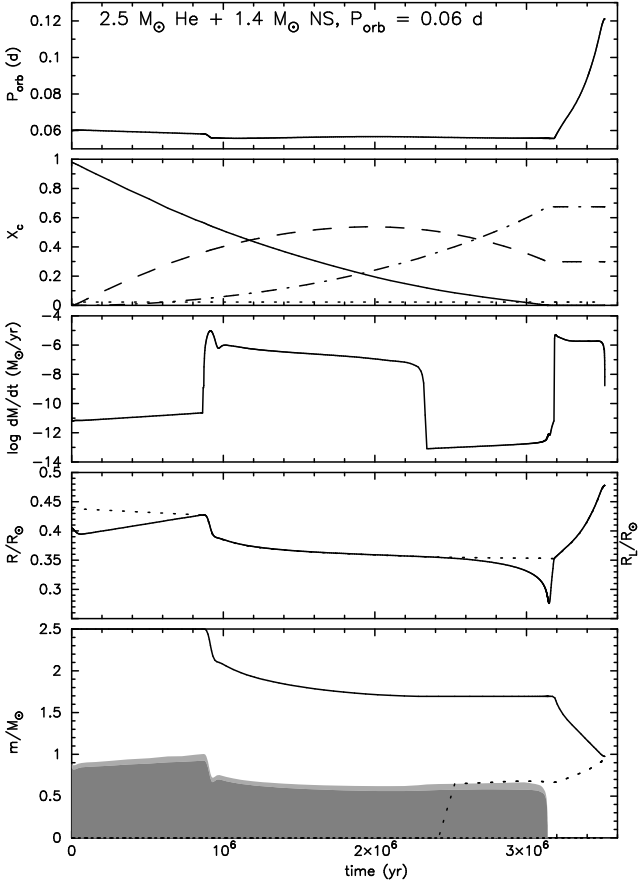,width=\linewidth}}
	 \caption[]{{\bf Case BA.} Same as Fig.~\ref{helium:fig:BA1p5} for a 
          2.5~\msun\ helium star. The peak and a dip in the mass-loss rate 
          mark the rapid, unstable mass-transfer phase, and the transition 
          to a slow, stable phase, respectively.}
	 \label{helium:fig:BA2p5}
	\end{figure}

RLOF is initially unstable and results in a rapid, thermal-timescale phase of mass transfer. The rapid loss of mass causes a rapid contraction of the star. The mass of the convective core decreases rapidly when the star adjusts to the new total mass.  The luminosity and effective temperature decrease, as well as the central temperature and the orbital period, until the mass-loss rate reaches a maximum value. The mass of the convective core then decreases slowly and the star contracts slowly. Mass transfer continues stably on a nuclear timescale. In a 2.5~\msun\ helium star, the transition between the rapid, thermal-timescale phase and the slow, stable mass-transfer phase is marked by a minimum in mass-loss rate (see Fig.~\ref{helium:fig:BA2p5}). The mass of the convective core reaches a minimum value before it increases and then decreases slowly. The further evolution is very similar to that of the lower-mass stars described in Sect.~\ref{helium:subsec:BA1p5}.

As in the case of 1.5 -- 2.1~\msun\ helium stars, the star shrinks inside its Roche lobe before helium is depleted in the centre. The star again fills its Roche lobe when the radius increases after helium is exhausted in the centre (case BAB).

\subsection{Roche-lobe overflow from 2.8 -- 2.9 $\bmath{\mathrm{M_{\sun}}}$ 
helium stars}
\label{helium:subsec:BA2p8}

\begin{figure}
 	 \centerline{\psfig{file=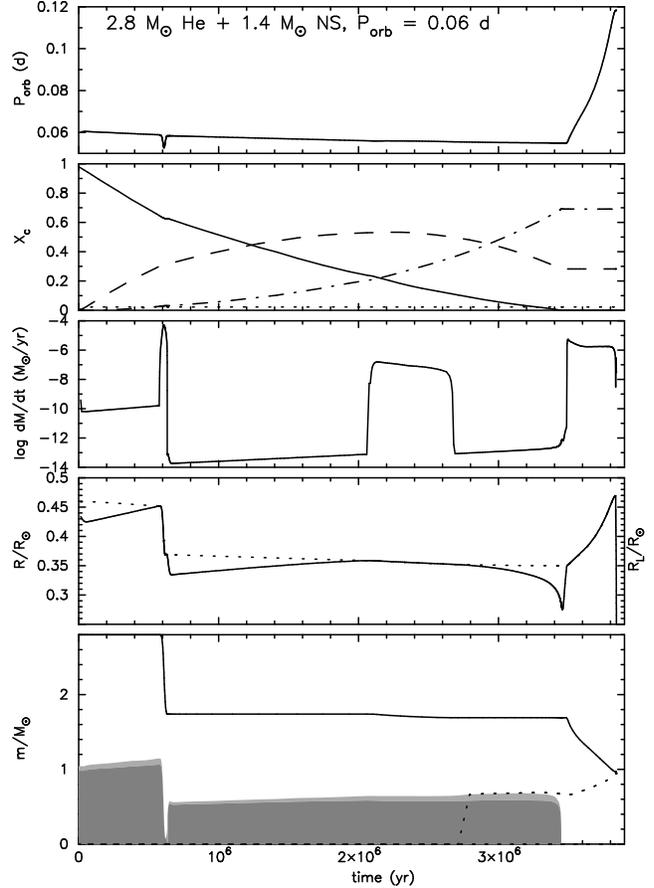,width=\linewidth}}
	 \caption[]{{\bf Case BA.} Same as Fig.~\ref{helium:fig:BA1p5} for a 
          2.8~\msun\ helium star. Note the constancy in central abundance, a 
          dip in the period, and the disappearance of the convective core 
          during the first, unstable mass-transfer phase.}
	 \label{helium:fig:BA2p8}
	\end{figure}

The mass transfer from 2.8 -- 2.9~\msun\ helium stars is also initiated (point 1 in Fig.~\ref{helium:fig:central}) by a thermal-timescale phase. This time, the rapid mass transfer with $q \sim 2$ causes the star to lose about 40 per cent of its mass rapidly. The central temperature decreases below the required temperature for helium burning so that the energy production in the centre ceases (point 2, see also Fig.~\ref{helium:fig:BA2p8}). This, in turn, causes rapid contraction of the star and disappearance of the convective helium core. The period decreases rapidly. After the maximum mass-loss rate is reached, the star contracts inside its Roche lobe, the orbit widens due to the loss of mass from the system, and the system is detached (point 3).

After thermal stability is restored (point 4), the convective core reappears, helium is reignited, and the star slowly expands. Wind mass loss is now that of a $\sim$ 1.7 \msun\ helium star. The orbital period again decreases due to gravitational-wave radiation. In a 2.8~\msun\ helium star in a very close orbit, at the moment when the central carbon abundance reaches a maximum, the stellar radius reaches its Roche radius, and a brief phase of stable RLOF takes place (point 5 to 6). Binaries of the same mass in a slightly wider orbit (as well as those with $M_{\mathrm{He}}$ = 2.9~\msun) do not undergo this second, slow phase of mass transfer since the maximum radius (which relates to the maximum energy-production rate and to the maximum central carbon abundance) is still within its Roche radius. However, the net effect is the same since during the slow phase only an insignificant amount (2 per cent) of mass is removed from the helium star. As in the case of lower-mass stars, the 2.8 -- 2.9~\msun\ helium stars also undergo case BAB mass transfer (point 7 to 8) with a lower mass-loss rate than that during the first, rapid RLOF phase.

In more massive helium stars ($M_{\mathrm{He}} > 2.9$~\msun), case BA causes the loss of more than half of the stellar mass. The rapid mass loss causes the disappearance of the convective helium core. However, the star is unable to recover its thermal equilibrium. RLOF becomes dynamically unstable and the binary will probably coalesce. 

\subsection{The role of gravitational-wave radiation}
\label{helium:subsec:grav-wave}

Before the star fills its Roche lobe, the orbit is tightened due to gravitational-wave radiation. In order to see the effect of gravitational-wave radiation, we also studied the case in which gravitational-wave emission is turned off. The exclusion of loss of angular momentum due to gravitational-wave radiation will only delay and shorten the first RLOF phase. In this case, during the slow stable RLOF phase, the orbit expands due to the loss of mass from the system, and there is no maximum period at the moment the central carbon abundance reaches its maximum.

The exclusion of gravitational-wave emission in the case of higher-mass helium stars ($M_{\mathrm{He}} > 2.9$~\msun) with very close orbits allows the wind mass loss to govern the loss of angular momentum. The orbit is widened such that mass transfer will only take place after the exhaustion of helium in the centre (case BB).


\section{Results: Case BB Mass Transfer}
\label{helium:sec:BB}

We devide case BB mass transfer into two mass groups, i.e. (i) 1.5 -- 2.1~\msun\ helium stars, in which RLOF is terminated when the degeneracy border is crossed and a heavy, degenerate CO or ONe white dwarf is produced, and (ii) 2.4 -- 6.6~\msun\ helium stars, in which an ONe white dwarf or a neutron star is the outcome.


\subsection{Roche-lobe overflow from 1.5 -- 2.1 $\bmath{\mathrm{M_{\sun}}}$ 
helium stars}
\label{helium:subsec:lowmass}

Case BB mass transfer from 1.5 -- 2.1~\msun\ helium stars resembles case BAB in the same range of mass. The exhaustion of helium in the core causes the star to expand and fill its Roche lobe. After the degeneracy boundary is crossed, it contracts rapidly inside its Roche lobe, and the system is detached. The mass-loss rate increases while the duration of RLOF decreases with increasing initial period and initial mass ratio.

After the core becomes degenerate but before the system detaches, carbon burning occurs in the radiative shell with maximum temperature. In 2.1~\msun\ helium stars, off-centre C-ignition takes place in the shell with maximum temperature. At first, carbon burns in a broad radiative shell (and marginally in the centre). When the off-centre convective shell appears, the whole core expands and the region interior to the convective shell cools. C-burning stops completely in the inner core, while convective off-centre burning continues. The calculations break down shortly after this point. It is expected (see Pols 2002 for a discussion) that eventually C-burning reaches the centre and a degenerate ONe core is formed. In the 1.5 and 1.8~\msun\ helium stars, the temperature is never high enough to cause significant carbon burning.

Before RLOF occurs, the gravitational-wave radiation brings the components closer together. In the systems with wider orbits ($P \geq 0\fd5$), the effect of gravitational-wave radiation is weak so that the period remains constant. The evolution of the period is as expected from Sect.~\ref{helium:subsec:orbital}.

We are not always able to follow the evolution until the detachment. However, the drop in mass-loss rate suggests that RLOF is about to end. Almost the entire envelope is removed from the star during RLOF, supporting Delgado \& Thomas' (1981) result. The remnants of 1.5 and 1.8~\msun\ helium stars are degenerate CO white dwarfs with masses of 0.9 -- 1.05~\msun, which are more massive than the remnants of case BAB mass transfer from helium stars in the same range of mass. This trend resembles Pols' (1994) result for case AB and B mass transfers. Helium stars with initial mass 2.1~\msun\ produce ONe white dwarfs.

\setlength{\tabcolsep}{2.0pt}
	\begin{table}  
         \caption[]{The binary parameters in case BB mass transfer from 1.5 -- 
          2.1~\msun\ helium stars: the initial mass and period; the duration, 
          amount of mass removed from the helium star, and the maximum mass-loss 
          rate during the Roche-lobe overflow; the final mass and period, the 
          final mass of the helium in the envelope and the final mass of the 
          neutron star. Masses are in solar mass, mass-loss rate in \rate, time 
          in yr, and period in days.}
	 \label{helium:tab:BBlowmass}
	 \begin{center}
	 \begin{tabular}{rr|rrr|rrrr}
	  \hline
	  \hline
          \noalign{\smallskip}
          $M_{\mathrm{i}}$ & $P_{\mathrm{i}}$~ & $\Delta t_{\mathrm{RLOF}}$~~ & 
          $\Delta M$ & $\dot{M}_{\mathrm{max}}$~~~ & $M_{\mathrm{f}}$~ & 
          $P_{\mathrm{f}}$~ & $M_{\mathrm{He,e}}$ & $M_{\mathrm{NS,f}}$ \\
          \noalign{\smallskip}
	  \hline
          \noalign{\smallskip}
          1.5 & 0.08 & \xpn{3.81}{5} & 0.581 & \xpn{2.2}{-6} & 
                0.897 & 0.16 & 0.0154 & 1.411 \\
              & 0.50 & \xpn{1.16}{5} & 0.559 & \xpn{9.8}{-6} & 
                0.940 & 0.97 & 0.0168 & 1.404 \\
              & 1.00 & \xpn{8.77}{4} & 0.527 & \xpn{1.5}{-5} & 
                0.946 & 1.90 & 0.0164 & 1.403 \\
          1.8 & 0.08 & \xpn{2.41}{5} & 0.793 & \xpn{5.6}{-6} & 
                0.996 & 0.16 & 0.0098 & 1.408 \\
              & 0.50 & \xpn{6.53}{4} & 0.749 & \xpn{2.4}{-5} & 
                1.043 & 0.91 & 0.0070 & 1.402 \\
              & 1.00 & \xpn{4.94}{4} & 0.748 & \xpn{3.8}{-5} & 
                1.046 & 1.79 & 0.0067 & 1.402 \\
          2.1 & 0.08 & \xpn{1.51}{5} & 0.966 & \xpn{1.2}{-5} & 
                1.126 & 0.14 & 0.0220 & 1.404 \\
              & 0.50 & \xpn{3.22}{4} & 0.874 & \xpn{4.8}{-5} & 
                1.224 & 0.73 & 0.0567 & 1.401 \\
              & 1.00 & \xpn{2.03}{4} & 0.852 & \xpn{7.6}{-5} & 
                1.233 & 1.43 & 0.0632 & 1.401 \\
          \noalign{\smallskip}
	  \hline
	  \hline
         \end{tabular}
	 \end{center}
	\end{table}
\setlength{\tabcolsep}{6pt}


\subsection{Roche-lobe overflow from 2.4 - 6.6 $\bmath{\mathrm{M_{\sun}}}$ 
helium stars}
\label{helium:subsec:highmass}

\begin{figure*}
 	 \centerline{\psfig{file=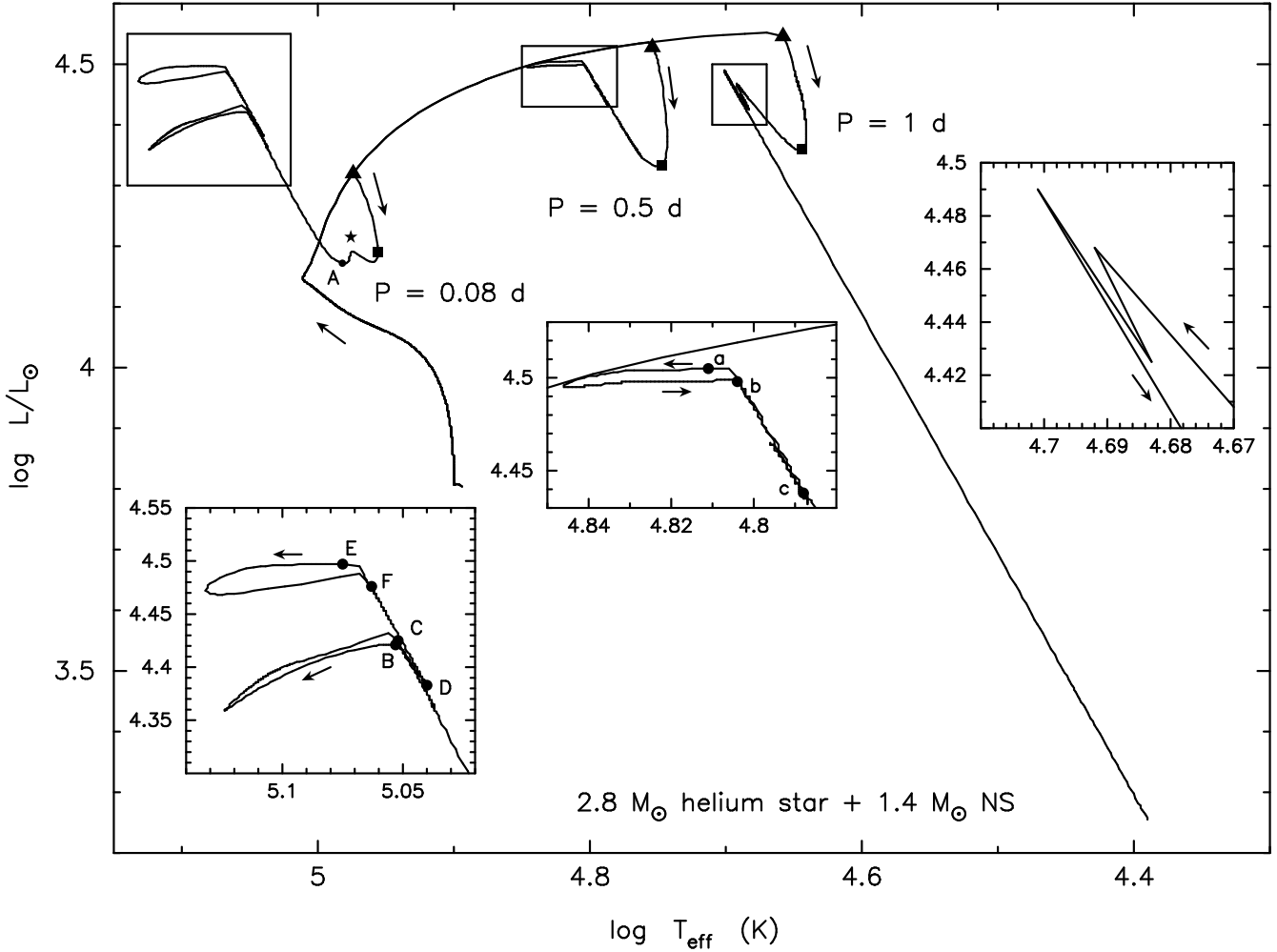,angle=270.,width=\linewidth}}
	 \caption[]{The HR diagram for a 2.8~\msun\ helium star with a 
	  1.4~\msun\ neutron-star companion in a close ($P_{\mathrm{i}} = 
	  0\fd08$), intermediate ($P_{\mathrm{i}} = 0\fd5$), and wide orbit 
          ($P_{\mathrm{i}} = 1^{\mathrm{d}}$). The triangles mark the onset of 
          (first) mass transfer phase, and the rectangles the (first) maximum 
          mass-loss rate. See text for an explanation about the labels.}
	 \label{helium:fig:hrd2p8}
	\end{figure*}

The following classification is based on the time of occurrence of the maximum mass-transfer rate; whether it occurs far before, just before, or during carbon core burning. This can be easily recognized from their tracks in the HR diagram, as are shown in Fig.~\ref{helium:fig:hrd2p8}. The close orbit systems are characterized by a small hump (indicated by a star) which marks the transition from the rapid to the slow mass-transfer phase. The wide orbit systems are characterized by the up and down movement (see the right-hand side inset in Fig.~\ref{helium:fig:hrd2p8}) which is caused by the decreasing and increasing of the mass-loss rate due to the appearance and disappearance of the C-burning convective shells (except in a 2.4~\msun\ helium star). The intermediate orbit systems show none of the two characteristics.

Before RLOF is initiated, gravitational-wave radiation only plays a role in shrinking the orbit in helium stars with masses of 2.4 -- 3.4~\msun\ in close orbits during the early phase of their evolution. In stars of the same masses in wider orbits, gravitational-wave emission is negligible so that the orbit remains constant. Wind mass loss dominates the loss of angular momentum and widens the orbit during the giant phase of the above range of mass, and during the whole evolution of $M_{\mathrm{He}} \geq 3.7$~\msun\ stars.

	\begin{table*}  
         \caption[]{Same as Table~\ref{helium:tab:BBlowmass} for 2.4 -- 
          6.6~\msun\ helium stars.$^{*}$ and $^{**}$ denote the second and third 
          phases of Roche-lobe overflow, respectively. The number in the {\it 
          Note} column indicates the carbon shell burning in which the 
          calculation stops, {\it a} implies the appearance of the shell, {\it 
          b} the shell burning, and {\it d} the disappearance of the shell; {\it 
          c} signifies that the calculation goes up to the penetration of the 
          helium-burning convective shell to the surface. For 2.4 and 2.5~\msun\ 
          stars, {\it cb} denotes the C-burning and {\it cd} the depletion of 
          carbon in the off-centre shell.}
	 \label{helium:tab:BBhighmass}
	 \begin{center}
	 \begin{tabular}{r@{\ \ }rr@{\ \ }r@{\ \ }rrr@{\ \ }r@{\ \ }r@{\ \ }rl}
	  \hline
	  \hline
          \noalign{\smallskip}
          $M_{\mathrm{i}}$ & $P_{\mathrm{i}}$~ & \multicolumn {3}{c|}{Case BB} & 
          $\Delta t_{\mathrm{detached}}$ & $M_{\mathrm{f}}$~ & 
          $P_{\mathrm{f}}$~ & $M_{\mathrm{He,e}}$ & $M_{\mathrm{NS,f}}$ & 
          Note \\
          &  & $\Delta t_{\mathrm{RLOF}}$~~ & $\Delta M$ & 
          $\dot{M}_{\mathrm{max}}$~~~ & & & & & & \\
          \noalign{\smallskip}
	  \hline
          \noalign{\smallskip}
          2.4 & 0.08    & \xpn{9.98}{4} & 1.082 & \xpn{2.4}{-5} &               
              & 1.305 & 0.11 & 0.0469 & 1.403 & cd \\
              & 0.50    & \xpn{1.82}{4} & 0.931 & \xpn{8.4}{-5} &               
              & 1.443 & 0.58 & 0.1076 & 1.400 & cd \\
              & 2.00    & \xpn{6.18}{3} & 0.836 & \xpn{2.3}{-4} &               
              & 1.528 & 2.19 & 0.1864 & 1.400 & cd \\
          2.5 & 0.08    & \xpn{8.84}{4} & 1.125 & \xpn{3.1}{-5} &               
              & 1.374 & 0.10 & 0.0602 & 1.403 & cb \\
              & 0.50    & \xpn{1.31}{4} & 0.878 & \xpn{9.8}{-5} &               
              & 1.561 & 0.52 & 0.1675 & 1.400 & cb \\
              & 2.00    & \xpn{7.14}{3} & 0.773 & \xpn{1.9}{-4} &               
              & 1.675 & 2.02 & 0.2734 & 1.400 & cd \\
          \hline
          2.8 & 0.08    & \xpn{6.37}{4} & 1.222 & \xpn{5.4}{-5} & \xpn{4.68}{3} 
              &       &      &        &       & \\
              & $^{*}$  & \xpn{1.35}{3} & 0.042 & \xpn{5.3}{-5} & \xpn{9.02}{2}
              &       &      &        &       & \\
              & $^{**}$ & \xpn{4.14}{1} & 0.001 & \xpn{1.9}{-5} &
              & 1.528 & 0.08 & 0.0411 & 1.400 & 2b \\
              & 0.50    & \xpn{9.34}{3} & 0.929 & \xpn{1.7}{-4} & \xpn{2.08}{3}
              &       &      &        &       & \\
              & $^{*}$  & \xpn{2.40}{3} & 0.111 & \xpn{4.3}{-5} &
              & 1.686 & 0.47 & 0.1033 & 1.400 & 2b \\
              & 1.00    & \xpn{8.65}{3} & 1.017 & \xpn{2.4}{-4} &
              & 1.701 & 0.93 & 0.1071 & 1.400 & 4a,c \\
          2.9 & 0.08    & \xpn{5.85}{4} & 1.252 & \xpn{6.5}{-5} & \xpn{3.85}{3} 
              &       &      &        &       & \\
              & $^{*}$  & \xpn{1.39}{3} & 0.046 & \xpn{5.8}{-5} & \xpn{5.34}{2}
              &       &      &        &       & \\
              & $^{**}$ & \xpn{3.22}{1} & 0.001 & \xpn{1.6}{-6} &
              & 1.592 & 0.07 & 0.0453 & 1.400 & 2a \\
              & 0.50    & \xpn{8.35}{3} & 0.922 & \xpn{1.9}{-4} & \xpn{9.62}{2}
              &       &      &        &       & \\
              & $^{*}$  & \xpn{2.02}{3} & 0.107 & \xpn{9.6}{-5} &
              & 1.775 & 0.44 & 0.1259 & 1.400 & 2a \\
              & 1.00    & \xpn{6.72}{3} & 0.975 & \xpn{2.3}{-4} &
              & 1.834 & 0.87 & 0.1726 & 1.400 & 3a,c \\
          3.1 & 0.08    & \xpn{4.79}{4} & 1.307 & \xpn{9.0}{-5} & \xpn{2.81}{3}
              &       &      &        &       & \\
              & $^{*}$  & \xpn{1.39}{3} & 0.054 & \xpn{6.6}{-5} &
              & 1.716 & 0.06 & 0.0487 & 1.400 & 2b \\
              & 0.30    & \xpn{1.45}{4} & 1.123 & \xpn{1.5}{-4} &
              & 1.887 & 0.23 & 0.1113 & 1.400 & 3b,c \\
              & 0.50    & \xpn{8.20}{3} & 1.015 & \xpn{2.6}{-4} &
              & 1.981 & 0.40 & 0.1979 & 1.400 & 2b \\
          3.2 & 0.08    & \xpn{4.39}{4} & 1.328 & \xpn{1.1}{-4} & \xpn{2.33}{3}
              &       &      &        &       & \\
              & $^{*}$  & \xpn{1.21}{3} & 0.053 & \xpn{6.6}{-5} &
              & 1.786 & 0.06 & 0.0528 & 1.400 & 2b \\
              & 0.30    & \xpn{1.23}{4} & 1.095 & \xpn{1.6}{-4} &
              & 1.976 & 0.22 & 0.1284 & 1.400 & 3b,c \\
              & 0.50    & \xpn{6.96}{3} & 1.003 & \xpn{2.8}{-4} &
              & 2.079 & 0.38 & 0.2281 & 1.400 & 2b \\
          3.4 & 0.08    & \xpn{3.70}{4} & 1.369 & \xpn{1.3}{-4} & \xpn{1.66}{3}
              &       &      &        &       & \\
              & $^{*}$  & \xpn{8.90}{2} & 0.044 & \xpn{6.2}{-5} &
              & 1.919 & 0.05 & 0.0635 & 1.400 & 2b \\
              & 0.30    & \xpn{9.42}{3} & 1.083 & \xpn{2.2}{-4} &
              & 2.154 & 0.20 & 0.1876 & 1.400 & 2b \\
              & 0.40    & \xpn{6.55}{3} & 1.035 & \xpn{2.9}{-4} &
              & 2.206 & 0.28 & 0.2333 & 1.400 & 2a \\
          3.6 & 0.09    & \xpn{3.06}{4} & 1.363 & \xpn{1.6}{-4} & \xpn{1.10}{3}
              &       &      &        &       & \\
              & $^{*}$  & \xpn{8.30}{2} & 0.044 & \xpn{1.4}{-4} &
              & 2.059 & 0.06 & 0.0676 & 1.400 & 2d,c \\
              & 0.25    & \xpn{1.03}{4} & 1.141 & \xpn{2.2}{-4} &
              & 2.258 & 0.16 & 0.1874 & 1.400 & 2a \\
              & 0.40    & \xpn{5.53}{3} & 1.025 & \xpn{3.3}{-4} &
              & 2.367 & 0.27 & 0.2804 & 1.400 & 2d,c \\
          3.8 & 0.09    & \xpn{2.61}{4} & 1.358 & \xpn{1.8}{-4} & \xpn{8.82}{2}
              &       &      &        &       & \\
              & $^{*}$  & \xpn{6.62}{2} & 0.024 & \xpn{5.7}{-5} &
              & 2.183 & 0.06 & 0.0955 & 1.400 & 2a \\
              & 0.25    & \xpn{9.15}{3} & 1.148 & \xpn{2.7}{-4} &
              & 2.366 & 0.16 & 0.1884 & 1.400 & 3b,c \\ 
              & 0.30    & \xpn{6.27}{3} & 1.080 & \xpn{3.2}{-4} &
              & 2.399 & 0.20 & 0.2273 & 1.400 & 2a,c \\
          4.0 & 0.09    & \xpn{2.40}{4} & 1.347 & \xpn{2.0}{-4} & \xpn{7.20}{2}
              &       &      &        &       & \\
              & $^{*}$  & \xpn{6.20}{2} & 0.021 & \xpn{6.5}{-5} &
              & 2.268 & 0.06 & 0.1039 & 1.400 & 2b,c \\
              & 0.20    & \xpn{1.04}{4} & 1.173 & \xpn{2.5}{-4} &
              & 2.419 & 0.13 & 0.1917 & 1.400 & 2b \\
              & 0.40    & \xpn{3.91}{3} & 0.791 & \xpn{3.6}{-4} &
              & 2.776 & 0.30 & 0.5229 & 1.400 & 3b \\
          4.4 & 0.09    & \xpn{1.85}{4} & 1.293 & \xpn{1.7}{-4} &
              & 2.392 & 0.06 & 0.1199 & 1.400 & 2a,c \\
              & 0.15    & \xpn{1.03}{4} & 1.144 & \xpn{2.3}{-4} &
              & 2.516 & 0.11 & 0.1958 & 1.400 & 1b \\
              & 0.30    & \xpn{4.33}{3} & 0.864 & \xpn{3.5}{-4} &
              & 2.774 & 0.24 & 0.4233 & 1.400 & 2d \\
          4.8 & 0.08    & \xpn{1.56}{4} & 1.211 & \xpn{1.8}{-4} &
              & 2.481 & 0.07 & 0.1385 & 1.400 & 1b \\
              & 0.15    & \xpn{6.96}{3} & 1.061 & \xpn{2.8}{-4} &
              & 2.586 & 0.13 & 0.2038 & 1.400 & 1a \\
              & 0.30    & \xpn{3.32}{3} & 0.552 & \xpn{3.7}{-4} &
              & 3.092 & 0.33 & 0.6739 & 1.400 & 2d \\
          5.1 & 0.08    & \xpn{1.17}{4} & 1.110 & \xpn{2.0}{-4} &
              & 2.555 & 0.08 & 0.1505 & 1.400 & 1b \\
              & 0.10    & \xpn{8.94}{3} & 1.041 & \xpn{2.2}{-4} &
              & 2.599 & 0.10 & 0.1775 & 1.400 & 1b \\
              & 0.20    & \xpn{4.27}{3} & 0.841 & \xpn{3.4}{-4} &
              & 2.802 & 0.21 & 0.3377 & 1.400 & 2d \\
          5.4 & 0.08    & \xpn{1.07}{4} & 1.080 & \xpn{2.1}{-4} &
              & 2.591 & 0.09 & 0.1496 & 1.400 & 2b \\
              & 0.10    & \xpn{8.53}{3} & 1.027 & \xpn{2.3}{-4} &
              & 2.625 & 0.10 & 0.1696 & 1.400 & 2b \\
              & 0.20    & \xpn{3.18}{3} & 0.508 & \xpn{3.5}{-4} &
              & 3.201 & 0.27 & 0.5832 & 1.400 & 2b \\
          5.8 & 0.08    & \xpn{6.68}{3} & 1.003 & \xpn{2.7}{-4} &
              & 2.871 & 0.10 & 0.1641 & 1.400 & 2b,c \\
              & 0.10    & \xpn{4.89}{3} & 0.925 & \xpn{3.2}{-4} &
              & 2.930 & 0.12 & 0.2009 & 1.400 & 2b \\
          6.2 & 0.08    & \xpn{4.61}{3} & 0.886 & \xpn{3.4}{-4} &
              & 3.117 & 0.11 & 0.2183 & 1.400 & 2b \\
          6.6 & 0.08    & \xpn{3.36}{3} & 0.537 & \xpn{4.0}{-4} &
              & 3.601 & 0.15 & 0.4517 & 1.400 & 2b \\
          \noalign{\smallskip}
	  \hline
	  \hline
         \end{tabular}
	 \end{center}
	\end{table*}

\subsubsection{Close-orbit systems}
\label{helium:subsubsec:close}

In the close-orbit systems, mass transfer takes place initially on the donor star's thermal timescale. The star contracts and the luminosity and effective temperature -- as well as the period -- decrease, until the mass-loss rate reaches a maximum, \mext{max}{1}. \teff\ then increases until the point which marks the transition between the rapid- and slow-phase of mass transfer (the hump marked by a star in Fig.~\ref{helium:fig:hrd2p8}). $L$ reaches another minimum, while $P$ also decreases to a minimum (point A). The star then moves to the upper left in the HR diagram. 

The appearance of a C-burning convective core (or off-centre shell in $M_{\mathrm{He}} \leq 2.5$~\msun\ stars) causes an increase in mass-loss rate to another maximum, \mext{max}{2}, which varies from 33 (in a 5.4~\msun\ helium star) to 58 per cent (in a 2.4~\msun\ helium star) of \mext{max}{1}. After the convective core stops growing in mass in $M_{\mathrm{He}} \leq 4$~\msun\ stars, during carbon core burning, the mass-loss rate drops, the star moves to the left in the HR diagram and shrinks inside its Roche lobe (point B, see the left-hand side inset in Fig.~\ref{helium:fig:hrd2p8}). The period is relatively constant during this detached phase.

When the carbon abundance in the convective core drops below 0.1, the star expands again, moves to the right in HR diagram, and the orbit expands.\footnote{We could not follow the subsequent evolution in $M_{\mathrm{He}} \leq 2.5$~\msun\ stars. However, we expect that these stars will undergo the same phenomenon as is found in the more massive ones.} The second phase of RLOF occurs after the C-burning convective core has vanished (point C). The mass-loss rate increases rapidly and reaches a maximum, \mext{max}{3} (point D), when the first C-burning convective shell appears. When this shell disappears, the mass-loss rate drops again.

In $M_{\mathrm{He}} \leq 2.9$~\msun\ stars, the radius decreases and the components are again detached (point E). The star moves to the left again at constant period until it reaches a minimum radius when the second C-burning convective shell appears. The star fills its Roche lobe for the third time (point F) after the second convective shell has vanished. \mdot\ increases and reaches a maximum value, \mext{max}{4}. The second RLOF lasts shorter than the first one, and the third one shorter than the second, because the evolution of the interior region proceeds faster (Habets 1986a).

In 3.1 -- 4~\msun\ helium stars, RLOF takes place in two episodes. After the first C-burning convective shell disappears, the mass-loss rate decreases to a minimum, but RLOF continues without a detached phase. In 3.6 -- 4~\msun\ helium stars, the first shrinkage of the star inside its Roche lobe takes place later than in the lower-mass stars, i.e. when carbon is being depleted in the core. The radius reaches its minimum value after the core has vanished. With the appearance of the first C-burning convective shell, the star fills its Roche lobe for the second time.

In $M_{\mathrm{He}} > 4$~\msun\ stars, \mdot\ also decreases after reaching \mext{max}{2}, until carbon is exhausted in the core. The behaviour of \mdot\ subsequent to carbon core burning is erratic, but it never exceeds \mext{max}{1}. In this range of mass, the star never shrinks inside its Roche lobe.

Why detached phases occur in one case and not in the other, and why the detachments take place in different phases of carbon burning for different cases, can be understood as follows. The start of each convective carbon burning phase (both in the core and in subsequent shells) causes the core to expand. As a result, the envelope tends to contract and the star tends to shrink inside its Roche lobe. In between the convective phases, the core contracts and the envelope expands again. However, the reaction of the stellar surface is delayed by the thermal diffusion timescale between the core and the surface. This timescale is roughly constant for each star, and does not vary strongly with mass. On the other hand, the timescale for C-burning decreases strongly with each successive convective phase, as well as with increasing mass, and eventually become much shorter than the thermal timescale. As a result, the detachments are delayed or even absent for advanced phases and/or large masses.

The transition from rapid to slow mass transfer is not found in $M_{\mathrm{He}} > 5.4$~\msun\ stars. Stars in this mass range with periods shorter than those in Table~\ref{helium:tab:BBhighmass} go through dynamically unstable mass transfer.

	\begin{figure}
 	 \centerline{\psfig{file=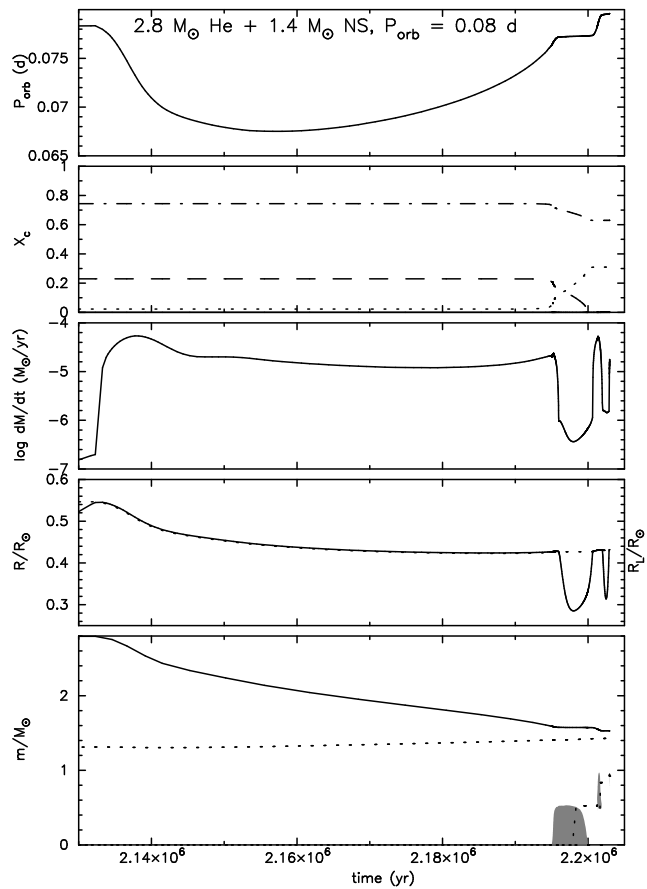,width=\linewidth}}
	 \caption[]{{\bf Case BB.} Same as Fig.~\ref{helium:fig:BA1p5} for a 
          2.8~\msun\ helium star with a 1.4~\msun\ neutron star in a close orbit, 
          during the case BB mass transfer. In the fifth panel, the upper and 
          lower dotted lines show the CO and O-Ne-Mg core masses (the latter is 
          defined as the mass where the carbon abundance is below 0.1); and the 
          shaded area mark the carbon-burning convective regions.}
	 \label{helium:fig:BBclose}
	\end{figure}

\subsubsection{Intermediate-orbit systems}
\label{helium:subsubsec:inter}

In the intermediate-orbit systems, RLOF occurs on the thermal timescale, the orbit shrinks, and the maximum mass-transfer rate, \mext{max}{1}, is reached not long before the appearance of the C-burning convective core (or off-centre convective shell in 2.4 -- 2.5~\msun\ helium stars). The star moves afterwards to the upper left in the HR diagram (see Fig.~\ref{helium:fig:hrd2p8}). Unlike in the close-orbit systems, the mass-transfer rate does not increase when the convective core appears. Along with the increase in $L$ and \teff, \mdot\ decreases.

\begin{figure}
 	 \centerline{\psfig{file=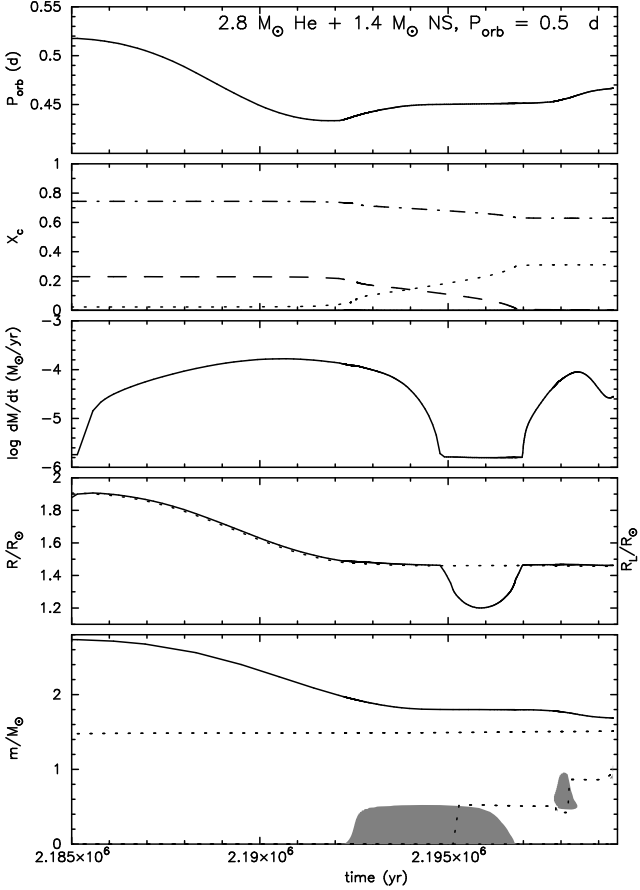,width=\linewidth}}
	 \caption[]{{\bf Case BB.} Same as Fig.~\ref{helium:fig:BBclose} for 
          a 2.8~\msun\ helium star in an intermediate orbit.}
	 \label{helium:fig:BBinter}
	\end{figure}

In 2.4 -- 2.9~\msun\ helium stars, \mdot\ decreases until the star shrinks inside its Roche lobe during carbon core (or off-centre shell) burning, and the system is detached (point a, see the middle inset of Fig.~\ref{helium:fig:hrd2p8} and Fig.~\ref{helium:fig:BBinter}). The period increases shortly after \mext{max}{1} is reached, and is relatively constant during the detached phase. When the carbon abundance in the core (or the off-centre shell) drops below 0.1, the star reaches the bluest point (In 2.4 -- 2.5~\msun\ helium stars, at this point the off-centre shell penetrates to the centre). The star expands, moves back to the right in the HR diagram, and fills its Roche lobe after the C-burning convective core has disappeared (point b). During the second RLOF, \teff\ and $L$ decrease while the period increases. After the first C-burning convective shell disappears, the mass-transfer rate reaches a maximum, \mext{max}{2} (point c), before decreasing again (see  Fig.~\ref{helium:fig:BBinter}).

In 3.1 -- 5.8~\msun\ helium stars in intermediate orbits, RLOF takes place without detached stages. After \mext{max}{1}, \mdot\ goes through a minimum and then increases again. Also due to the difference in thermal timescale between the core and the envelope, the occurence of the minimum \mdot\ (particularly in this range of mass) varies. Like in the close-orbit systems, the period decreases throughout the whole mass-transfer phase.

\subsubsection{Wide-orbit systems}
\label{helium:subsubsec:wide}

Thermal-timescale RLOF is most rapid in wide-orbit systems, and the maximum mass-loss rate, \mext{max}{1}, is reached during the carbon core (or off-centre shell) burning. The mass-transfer rate then decreases and reaches a minimum value (in a 2.4~\msun\ helium star, the decrease in \mdot\ leads to a detachment), before reaching another maximum, \mext{max}{2}. The increases and decreases in \mdot\ can be seen in the HR diagram as the up and down movements (see Fig.~\ref{helium:fig:hrd2p8}). Except in a 2.4~\msun\ helium star, RLOF in wide-orbit systems takes place without any detached stages. Again, only in $M_{\mathrm{He}} < 2.9$~\msun\ stars does the period reach a minimum value shortly after \mext{max}{1}.

       \begin{figure*}
 	 \centerline{\psfig{file=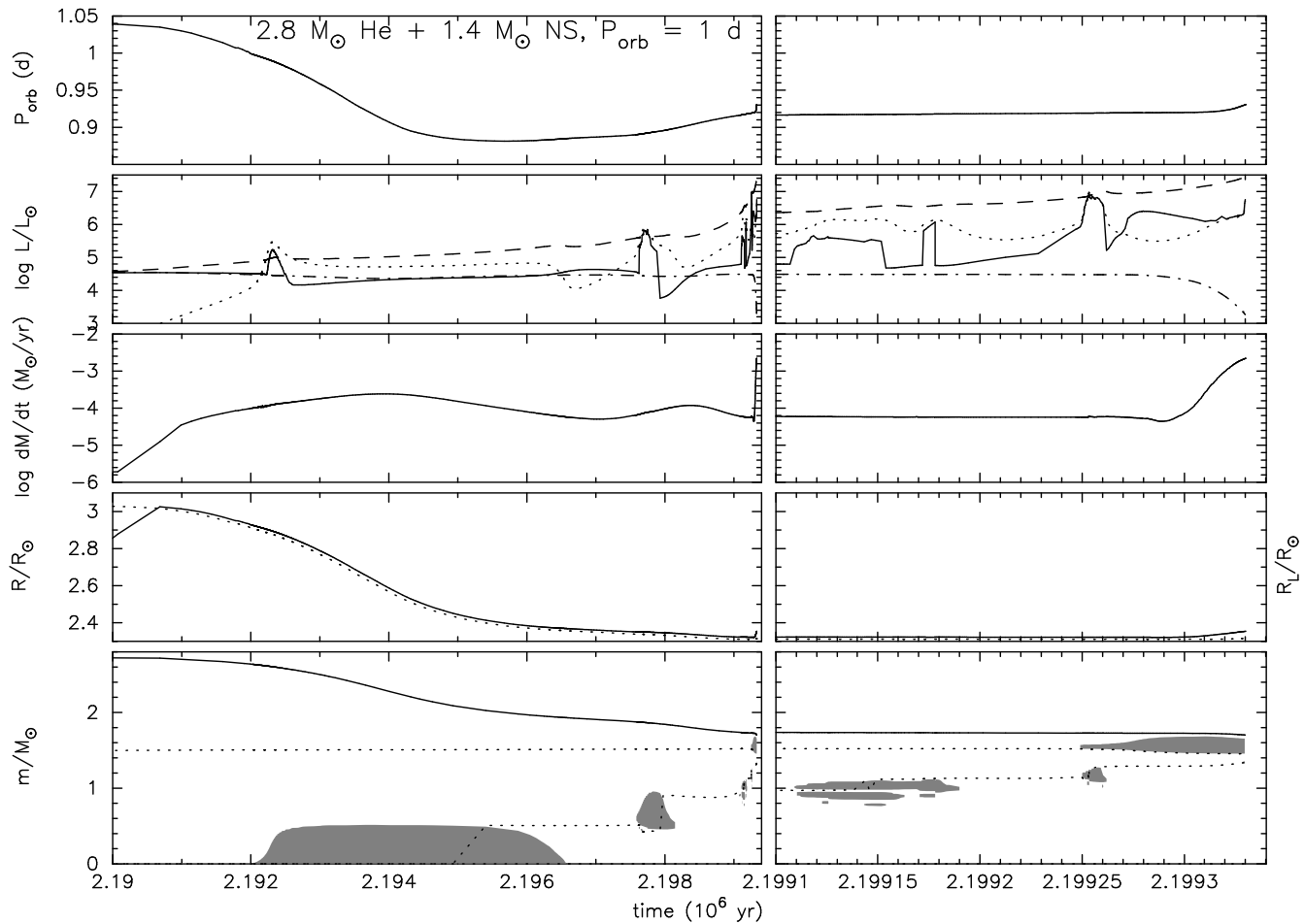,angle=270.,width=\linewidth}}
         \caption[]{{\bf Case BB.} Same as Fig.~\ref{helium:fig:BBclose} for 
          a 2.8~\msun\ helium star in a wide orbit. Instead of central abundances, 
          the second panels show the stellar luminosity (dash-dotted-), and the 
          contributions of helium burning (solid-), carbon burning (dotted-), 
          and neutrino losses (dashed-line). The panels on the right-hand side 
          show the evolution around the penetration of the He-burning convective 
          shell to the surface.}
         \label{helium:fig:BBwide}
        \end{figure*}

\subsubsection{The He-burning convective envelope and common-envelope evolution}
\label{helium:subsubsec:convective}

As can be seen in the right-hand panels of Fig.~\ref{helium:fig:BBwide}, a He-burning convective shell appears in the envelope before the appearance of the third C-burning convective shell. Such a shell also appears in previous helium star calculations (e.g. Habets 1986a, 1986b; Nomoto 1984, 1987; Pols 2002). Because of the small envelope mass, this shell almost penetrates to the surface after the disappearance of the third C-burning convective shell. In response to mass loss from a convective envelope, the star expands, \mdot\ increases enormously by 2 orders of magnitude, and the orbit widens. $L$ and \teff\ decrease and the star moves to the lower right in the HR diagram (see Fig.~\ref{helium:fig:hrd2p8}). In a 3.8~\msun\ helium star, the He-burning convective shell appears and penetrates to the surface before the second C-burning convective shell appears.

We are not always able to follow the evolution of the helium stars beyond the appearance of the second C-burning convective shell. However, we found that the behaviour occurs for those  masses and orbits that we were able to compute the furthest (see the {\it c} notation in Table~\ref{helium:tab:BBhighmass}) in helium stars less massive than 4~\msun. Therefore, we expect that this is a typical phenomenon of the late stage of a case BB mass transfer from $M_{\mathrm{He}} < 4$~\msun\ stars (in RLOF initiated during carbon shell burning (Habets 1986a), the envelope is thick so that the expansion of He-burning convective shell occurs deep inside the envelope). In $M_{\mathrm{He}} \geq 4$~\msun\ the start of a convective shell is only found in close orbits, but not in the wide orbits athough the calculations were done up to the same stage of evolution. However, we cannot exclude that such a convective shell does not eventually develop in these more massive stars as well.

The outcome of mass transfer in the presence of a convective helium envelope is not clear. Mass transfer might become dynamically unstable which would lead to a CE and spiral-in phase. Although we do not find a convective envelope in helium stars more massive than 4~\msun\ in wider orbits, the large mass ratio ($q \ga 2.6$ at the onset of RLOF) and the high mass-loss rate (\mdot~$> 10^{-4}$~\rate) suggest that a CE and spiral-in phase will very likely also take place. However, the high luminosity due to neutrino emission ($L_{\nu} > 10^{7}$ \lsun) suggests that the core will collapse within a few decades after the end of the calculation. The closeness to collapse is supported by the fact that Ne-burning has already started radiatively off-centre when the fourth C-burning convective shell appears. The explosion will probably be observed as a type Ic supernova. Nevertheless, should the two components undergo a spiral-in, a merger will not necessarily occur. A crude calculation of the gravitational binding energy of the convective envelope to the CO-core boundary of a 3.2~\msun\ helium star when the penetration takes place yields $\lambda$ = 0.123. With $\eta_{\mathrm{CE}} = 1$, $M_{\mathrm{He}}$ = 1.98~\msun, $M_{\mathrm{core}}$ = 1.62~\msun\ and $P_{\mathrm{orb}} = 0\fd22$, we obtain from eq.~(\ref{helium:eq:webbink}) a final separation of 0.0729~\rsun\ (period of $1\fm89$), which is still larger than the radius of the core (0.0275~\rsun). Without helium left in the envelope, the explosion will be observed as a type Ic supernova. The close post-SN separation subjects the two neutron stars to a strong loss of gravitational-wave radiation, causing the two compact stars to merge on a timescale of 2\,500 years, which may produce a gamma-ray burst.

\subsubsection{Final masses and outcome}
\label{helium:subsubsec:mass}

The final masses and periods are shown in Table~\ref{helium:tab:BBhighmass}. Here final corresponds to the last stage of our calculations, which varies from the appearance of the first C-burning convective shell to the penetration of the He-burning convective shell to the surface. We found that after the carbon convective core has disappeared, only a small amount of mass is transferred to the neutron star. The period alters insignificantly. Hence, the values in Table~\ref{helium:tab:BBhighmass} can be considered as the final masses and periods prior to SN explosion, unless a spiral-in phase occurs as discussed in Sect.~\ref{helium:subsubsec:convective}.

With increasing initial period and decreasing initial mass, the amount of transferred matter decreases. In 2.4 -- 2.5~\msun\ helium stars, the mass of the remnant ranges from 1.3 -- 1.6~\msun, but the mass of the CO core (1.21 -- 1.34~\msun) never exceeds the Chandrasekhar limiting mass, $M_{\mathrm{Ch}}$. These stars will develop degenerate ONe cores. The future of these systems is not clear since they could not be computed through carbon burning. If the ONe core can grow to 1.37 -- 1.39~\msun, they will undergo a SN explosion triggered by electron capture on Ne and Mg (e.g. Nomoto 1984, 1987), otherwise they will become heavy ONe white dwarfs.

Stars with $M_{\mathrm{He}} \ge 2.6 - 2.8$~\msun\ which undergo case BB mass transfer develop CO cores more massive than $M_{\mathrm{Ch}}$ (where the actual lower mass limit probably depends on the orbital period). These stars will eventually complete neon, oxygen and silicon burning and form iron cores. The collapse of the Fe core by photodisintegration triggers a SN explosion, leaving a neutron-star remnant. 

The final amount of helium left in the envelope probably determines whether the explosion will be a Type Ib or a Type Ic supernova. The main observational criterion to distinguish between these types is the detection of helium in the spectrum. Helium is generally absent in SNe Ic, although a small amount may be present (Filippenko et al. 1995). The maximum allowed amount of He is very uncertain, however, and may also depend on other factors such as mixing during the explosion (Woosley \& Eastman 1997).

For low-mass He stars and/or close orbits, we find there may be as little as 0.04~\msun\ of He left, and generally $<$ 0.10~\msun\ (see Table~\ref{helium:tab:BBhighmass}). This is less than is achieved by stellar-wind mass loss in massive He-star models (Woosley, Langer \& Weaver 1995; Wellstein \& Langer 1999; Pols 2002) and also less than obtained from conservative case ABB or case BB mass transfer models by Wellstein \& Langer (1999). We conclude that case BB mass transfer to a compact companion in a very-close binary is the most efficient way to produce almost bare CO cores prior to explosion. These stars will almost certainly produce a Type Ic supernova. Such a model was first suggested by Nomoto et al. (1994) as the progenitor of the Type Ic SN 1994I, and shown to match the observed lightcurve (Iwamoto et al. 1994).  The amount of He left in the envelope increases with increasing initial mass and initial period. Probably, the more massive case BB remnants and those from the widest orbits undergo a Type Ib explosion, as do helium stars that avoid case BB mass transfer (Pols 2002).


\subsection{The binary millisecond pulsars and double neutron-star pulsars}
\label{helium:subsec:pulsar}

\begin{figure}
 	 \centerline{\psfig{file=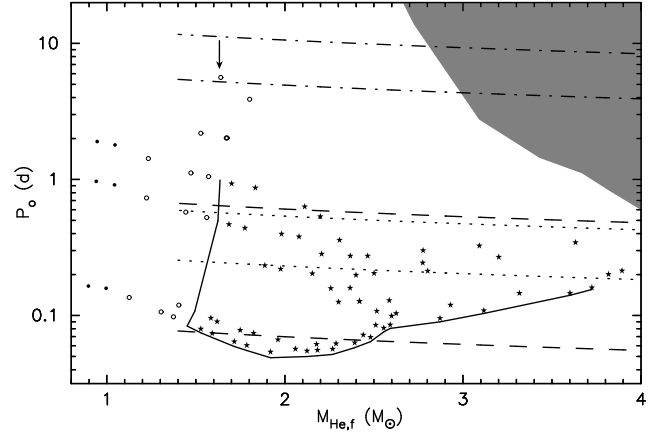,width=\linewidth}}
         \caption[]{Range of mass and period of the immediate pre-SN progenitor 
          of double neutron-star binaries: B1913+16 (dashed-), B1534+12 (dotted-), 
          and J1518+49 (dash-dotted-line). Upper and lower lines represent the 
          maximum and minimum pre-SN periods. The calculations done in this 
          work are marked according to the type of remnant they produce: CO 
          white dwarfs (solid-circle), ONe white dwarf (open-circle), or neutron 
          stars (solid-stars). The orbital periods ($P_{\mathrm{o}}$) and masses 
          $M_{\mathrm{He,f}}$ are those immediately prior to the SN or the formation 
          of the white dwarf.  The region below and to the left of the solid curve 
          is the zone of avoidance for the formation of neutron stars. The shaded 
          area marks the region where a double neutron star can be produced by 
          avoiding RLOF, taken from Pols' (2002) single helium stars calculation 
          after taking into account the effect of stellar wind mass loss.}
         \label{helium:fig:pulsar}
        \end{figure}

The outcome of case BB mass transfer from 1.5 -- 2.5~\msun\ helium stars are heavy white dwarfs with a neutron-star companion. Their positions in the period-mass plane (solid- and open-circles in Fig.~\ref{helium:fig:pulsar}) cover the region where the existence of binary millisecond pulsars with heavy white-dwarf companions cannot be explained by extreme mass transfer (Tauris, van den Heuvel \& Savonije 2000). Systems like J1756-5322 ($P_{\mathrm{orb}} = 0\fd453$, $M_{\mathrm{WD}} = 0.683$~\msun\ with the assumption that $i = 60^{\circ}$) are not produced by our nor Tauris et al.'s calculations. Therefore, they must have come from stars with $5 < M_{\mathrm{ZAMS}} / \mathrm{M_{\sun}} < 8$.

Case BB mass transfer from $M_{\mathrm{He}} \geq 2.6 - 2.8$~\msun\ produces neutron-star binaries. Attempts to find the progenitors of the three known galactic double neutron-star binaries (outside the globular cluster) have been carried out e.g. by Yamaoka, Shigeyama \& Nomoto (1993), Fryer \& Kalogera (1997), Bagot (1997), and recently Francischelli, Wijers \& Brown (2001). Their calculations were approximated by the evolution of single helium stars, and it is often assumed that the helium star does not fill its Roche lobe (i.e. $R_{\mathrm{He}} < R_{\mathrm{L,He}}$) in order to avoid an unstable phase of mass transfer. The orbital parameters of the observed binary pulsars are presented in Table~\ref{helium:tab:pulsar}. Here we will try to see if our helium star calculations can explain their existence.

	\begin{table}  
         \caption[]{The orbital parameters of double neutron-star pulsars: total 
          mass of the system in \msun, orbital period in days, and 
          eccentricity.}
	 \label{helium:tab:pulsar}
	 \begin{center}
	 \begin{tabular}{l@{\, \,}l@{\, \,}r@{\, \,}r@{\, \,}l}
	  \hline
	  \hline
          \noalign{\smallskip}
          PSR & ~$M_{\mathrm{T}}$ & $P_{\mathrm{b}}$~~ & $e$~~~ & Reference \\
          \noalign{\smallskip}
	  \hline
          \noalign{\smallskip}
          B1913+16 & 2.828 &  0.323 & 0.617 & Taylor \& Weisberg 1989\\
          B1534+12 & 2.679 &  0.421 & 0.274 & Wolszczan 1991\\
          J1518+49 & 2.62  &  8.634 & 0.249 & Nice, Sayer \& Taylor 1996\\
          \noalign{\smallskip}
	  \hline
	  \hline
         \end{tabular}
	 \end{center}
	\end{table}

If we simply assume that the radius of the pre-SN orbit (the helium-star binary), $a_{\mathrm{o}}$, must be inside the radius turning point of the post-SN orbit, $a_{\mathrm{p}}$, (Flannery \& van den Heuvel 1975) i.e.,
	\begin{eqnarray}
	(1-e) \, a_{\mathrm{p}} \leq a_{\mathrm{o}} \leq (1+e) \, a_{\mathrm{p}}
	\label{helium:eq:post-sep}	 
	\end{eqnarray}
where $e$ is the eccentricity of the binary pulsar, then from Kepler's third law we have the following condition
	\begin{eqnarray}
	(1-e)^{3} M_{\mathrm{p}} \, P_{\mathrm{p}}^{2} \leq 
	M_{\mathrm{o}} \, P_{\mathrm{o}}^{2} \leq 
	(1+e)^{3} M_{\mathrm{p}} \, P_{\mathrm{p}}^{2}
	\label{helium:eq:post-per}	 
	\end{eqnarray}
where $M_{\mathrm{o}}$, $M_{\mathrm{p}}$ are the total pre- and post-SN masses; $P_{\mathrm{o}}$, $P_{\mathrm{p}}$ are the pre- and post-SN periods.

We plotted the allowed minimum and maximum pre-SN periods according to eq.~(\ref{helium:eq:post-per}) for each double neutron-star binary in Fig.~\ref{helium:fig:pulsar}, under the assumption that the pulsar companion is a 1.4~\msun\ neutron star. Fig.~\ref{helium:fig:pulsar} shows that systems like B1913+16 are easy to produce from systems with initial masses of 2.4 -- 6.7~\msun\ in initial periods of $0\fd1$ -- $0\fd7$, while systems like B1534+12 can be made from helium stars with the same range of mass in a narrower range of period ($0\fd3$ -- $0\fd6$). None of the case BB models we have computed can produce B1518+49-like systems. The remnant of the 2.4~\msun\ system (indicated by the arrow in Fig.~\ref{helium:fig:pulsar}) is a ONe white dwarf/neutron star binary in a circular orbit. A case BC evolution might be able to explain the formation of this system, or a $M_{\mathrm{He}} \ga 3$~\msun\ star, which avoids RLOF altogether (the shaded area in Fig.~\ref{helium:fig:pulsar}).

The region to the left of the solid line in Fig.~\ref{helium:fig:pulsar} produces white dwarfs. The area below the solid curve will undergo case BA or dynamically unstable mass transfer (see Sect.~\ref{helium:sec:BA}). Therefore, these regions can be considered as the zone of avoidance for the formation of double neutron stars.

\section{Conclusions}
\label{helium:sec:conclusion}

        \begin{figure}
 	 \centerline{\psfig{file=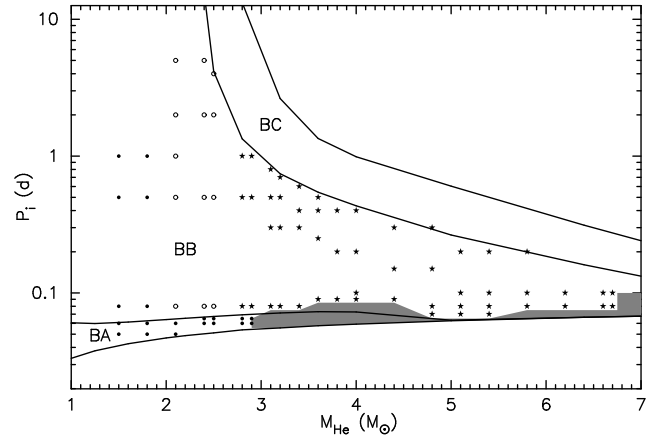,width=\linewidth}}
         \caption[]{The final fate of the helium stars borm in systems consisting of 
          helium star and neutron star: CO white dwarf (solid-circle), ONe white 
          dwarf (open-circle), or neutron star (solid-star). $P_{\mathrm{i}}$ and 
          $M_{\mathrm{He}}$ are the orbital period and helium star mass at the onset 
          of the evolution of the binary. The lines separate the regions of case BA, 
          BB, and BC mass transfers, taken from Pols' (2002) single helium stars 
          calculations. The shaded area mark the region where mass transfer is 
          dynamically unstable.}
         \label{helium:fig:final}
        \end{figure}

We have done calculations of helium stars with masses in the interval of 1.5 -- 6.7~\msun\ with a 1.4~\msun\ neutron star companion. We allowed the helium star to fill its Roche lobe during helium core burning (case BA) and helium shell burning (case BB mass transfer). The final fate of the helium star as a function of its initial mass and period is summarized in Fig.~\ref{helium:fig:final}.

Case BA mass transfer from helium stars with masses of 1.5 -- 2.9~\msun\ occurs during helium core burning and stops when the central abundance of helium drops below 0.1 and the star contracts. In helium stars less massive than 2.4~\msun, RLOF is stable and takes place on the nuclear timescale. For $2.4 \leq M_{\mathrm{He}} / \mathrm{M_{\sun}} \leq 2.9$, a rapid, thermal-timescale phase of RLOF is followed by a stable, slow phase of mass transfer on the nuclear timescale. Driven by radius expansion, a stable phase of case BAB mass transfer occurs during helium shell burning. This phase of mass transfer stops when the degeneracy border is crossed and the star contracts. During RLOF, the entire envelope is removed from the helium star. The remnants of case BA mass transfer are heavy, degenerate CO white dwarfs. This implies that a star with initial mass as large as 12~\msun\ still can become a white dwarf, as also found by e.g. Wellstein et al. (2001) in stars that go through conservative contact-free case A evolution.

Stable case BB mass transfer from 1.5 -- 2.1~\msun\ stars takes place on the nuclear timescale and stops when the degeneracy border is crossed and the star contracts. The entire envelope is removed during RLOF, and heavy, CO (from 1.5 -- 1.8~\msun) or ONe (from 2.1~\msun\ helium stars) white dwarfs are the remnants in this range of mass. Case BB mass transfer from 2.4 -- 2.5~\msun\ helium stars produces ONe white dwarfs and that from more massive ones produces neutron stars. RLOF from $M_{\mathrm{He}} \geq 2.4$~\msun\ takes place on the thermal timescale. In close-orbit systems, this is followed by a slow, stable phase of mass transfer. There is a tendency that the helium envelope becomes convective in helium stars less massive than 4~\msun. Mass transfer might become dynamically unstable in this case. However, the star probably collapses before a spiral-in can occur. Even if a CE and spiral-in phase does occur, the system is likely to survive. In both cases, a double neutron-star binary is formed.

From the amount of helium left in the envelope, we suggest that helium stars of high mass or in a wide orbit produce type Ib SNe. Type Ic SNe come from helium stars of lower mass or in a close orbit. We are also able to provide a zone of avoidance for the formation of double neutron stars.

\section*{Acknowledgements}
This work was sponsored by NWO Spinoza Grant 08-0 to E.~P.~J. van den Heuvel. It is a pleasure to thank Norbert Langer, Gijs Nelemans, Winardi Sutantyo, Thomas Tauris, and Lev Yungelson whose valuable comments and suggestions improve the early version of the manuscript; and Philipp Podsiadlowski for his constructive referee comments.

\label{lastpage}


\begin{thebibliography}{}
  \bibitem{} Avila Reese V. A., 1993, Rev. Mex. Astron. Astrofis. 25, 79
  \bibitem{} Bagot P., 1997, A\&A 322, 533
  \bibitem{} Bhattacharya D., van den Heuvel E. P. J., 1991, Phys. Rep. 203, 1
  \bibitem{} Brown G. E., 1995, ApJ 440, 270
  \bibitem{} Chevalier R. A., 1993, ApJ 411, L33
  \bibitem{} Chevalier R. A., 1996, ApJ 459, 322
  \bibitem{} Delgado A. J., Thomas H.-C., 1981, A\&A 96, 142
  \bibitem{} Dewi J. D. M., Tauris T. M., 2000, A\&A 360, 1043
  \bibitem{} de Jager C., Nieuwenhuijzen H., van der Hucht K. A., 1988, 
             A\&AS 72, 259
  \bibitem{} de Kool M., 1990, ApJ 358, 189
  \bibitem{} Eggleton P. P., 1971, MNRAS 151, 351
  \bibitem{} Eggleton P. P., 1972, MNRAS 156, 361
  \bibitem{} Eggleton P. P., 1973, MNRAS 163, 279
  \bibitem{} Eggelton P. P., 1983, ApJ 268, 368
  \bibitem{} Ergma E. V., Fedorova A. V., 1990, Ap\&SS 163, 143
  \bibitem{} Filippenko A. V. et al., 1995, ApJ 450, L11
  \bibitem{} Flannery B. P., van den Heuvel E. P. J., 1975, A\&A 39, 61
  \bibitem{} Francischelli G. J., Wijers R. A. M. J., Brown G. E., 2001, 
             submitted to ApJ, astro-ph/0103216
  \bibitem{} Fryer C., Kalogera V., 1997, ApJ 489, 244
  \bibitem{} Habets G. M. H. J., 1986a, A\&A 165, 95
  \bibitem{} Habets G. M. H. J., 1986b, A\&A 167, 61
  \bibitem{} Hamann W.-R., Koesterke L., Wessolowski U., 1995, A\&A 299, 151
  \bibitem{} Hamann W.-R., Sch\"{o}nberner D., Heber U., 1982, A\&A 116, 273
  \bibitem{} Iwamoto K., Nomoto K., Hoflich P., Yamaoka H., Kumagai S., 
             Shigeyama T., 1994, ApJ 437, L115
  \bibitem{} Landau  L. D., Lifshitz E., 1958, The classical theory of fields, 
             Pergamon Press, Oxford
  \bibitem{} Nelemans G., van den Heuvel E. P. J., 2001, A\&A 376, 950  
  \bibitem{} Nice D. J., Sayer R. W., Taylor J. H., 1996, ApJ 466, L87        
  \bibitem{} Nieuwenhuijzen H., de Jager C., 1990, A\&A 231, 134
  \bibitem{} Nomoto K., 1984, ApJ 277, 791
  \bibitem{} Nomoto K., 1987, ApJ 322, 206
  \bibitem{} Nomoto K., Yamaoka H., Pols O. R., van den Heuvel E. P. J., 
             Iwamoto K., Kumagai S., Shigeyama T., 1994, Nature 371, 227
  \bibitem{} Nugis T., Lamers H. J. G. L. M., 2000, A\&A 360, 227
  \bibitem{} Pols O. R. 1994, A\&A 290, 119          
  \bibitem{} Pols O. R. 2002, in preparation          
  \bibitem{} Pols O. R., Schr\"{o}der K.-P., Hurley J. R., Tout C. A., 
             Eggleton P. P. , 1998, MNRAS 298, 525
  \bibitem{} Pols O. R., Tout C. A., Eggleton P. P., Han Z., 1995, 
             MNRAS 274, 964
  \bibitem{} Savonije G. J., de Kool M., van den Heuvel E. P. J., 1986, 
             A\&A 155, 51
  \bibitem{} Soberman G. E., Phinney E. S., van den Heuvel E. P. J., 1997, 
             A\&A 327, 620
  \bibitem{} Taam R. E., King A. R., Ritter H., 2000, ApJ 541, 329
  \bibitem{} Tauris T. M., Dewi J. D. M., 2001, A\&A 369, 170 
  \bibitem{} Tauris T. M., Savonije G. J., 1999, A\&A 350, 928
  \bibitem{} Tauris T. M., van den Heuvel E. P. J., Savonije G. J., 2000, 
             ApJ 530, L93
  \bibitem{} Taylor J. H., Weisberg J. M., 1989, ApJ 345, 434
  \bibitem{} Tutukov A. V., Fedorova A. V., 1990, Sov. Astron. 33, 606
  \bibitem{} Tutukov A. V., Yungelson L., 1973, N. Info. 27, 70
  \bibitem{} van den Heuvel E. P. J., 1994, in Shore S. N., Livio M., 
             van den Heuvel E. P. J., eds., Interacting Binaries, Springer, 
             Berlin  
  \bibitem{} van Kerkwijk M. H., Charles P. A., Geballe T. R., et al., 1992, 
             Nature 355, 703  
  \bibitem{} van Kerkwijk M. H., Geballe T. R., King D. L., van der Klis M., 
             van Paradijs J., 1996, A\&A 314, 521     
  \bibitem{} Webbink R. F., 1984, ApJ 277, 355
  \bibitem{} Wellstein S., Langer N., 1999, A\&A 350, 148
  \bibitem{} Wellstein S., Langer N., Braun H., 2001, A\&A 369, 939
  \bibitem{} Wolszczan A., 1991, Nature 350, 688
  \bibitem{} Woosley S. E., Eastman R. G., 1997, in Ruiz-Lapuente P., Canal R., 
             Isern J., eds., Thermonuclear Supernovae, NATO ASIC Proc. 486, p. 821
  \bibitem{} Woosley S. E., Langer N., Weaver T. A., 1995, ApJ 448, 315
  \bibitem{} Yamaoka H., Shigeyama T., Nomoto K., 1993, A\&A 267, 433
\end{thebibliography}
\end{document}